# Where is the mind within the brain? Transient selection of subnetworks by metabotropic receptors and G protein-gated ion channels

Danko Nikolić*

*Frankfurt Institute for Advanced Studies
* Department of Psychiatry, Psychosomatic Medicine and Psychotherapy, University Hospital Frankfurt
*evocenta GmbH
*Robots Go Mental UG

Correspondence:

Dr. Danko Nikolić
Frankfurt Institute for Advanced Studies
Ruth-Moufang-Straße 1,
60438 Frankfurt am Main
Germany
danko.nikolic@gmail.com

**Abstract**

Perhaps the most important question posed by brain research is: How the brain gives rise to the mind. To answer this question, we have primarily relied on the connectionist paradigm: The brain's entire knowledge and thinking skills are thought to be stored in the connections; and the mental operations are executed by network computations. I propose here an alternative paradigm: Our knowledge and skills are stored in metabotropic receptors (MRs) and the G protein-gated ion channels (GPGICs). Here, mental operations are assumed to be executed by the functions of MRs and GPGICs. As GPGICs have the capacity to close or open branches of dendritic trees and axon terminals, their states transiently re-route neural activity throughout the nervous system. First, MRs detect ligands that signal the need to activate GPGICs. Next, GPGICs transiently selects a subnetwork within the brain. The process of selecting this new subnetwork is what constitutes a mental operation – be it in a form of directed attention, perception or making a decision. Synaptic connections and network computations play only a secondary role, supporting MRs and GPGICs. According to this new paradigm, the mind emerges within the brain as the function of MRs and GPGICs whose primary function is to continually select the pathways over which neural activity will be allowed to pass. It is argued that MRs and GPGICs solve the scaling problem of intelligence from which the connectionism paradigm suffers.

## 1. Introduction

Today, there is practically a unanimous agreement that the computations of the brain are achieved through network connectivity [1]. A variety of theories exist (e.g., [2, 3, 4]) but they all propose different ways of how network computations may take place; all these theories agree that the synaptic connectivity contains the bulk the knowledge learned by the brain. The existing theories also agree that central to mental operations are the voltages forming across neuron membranes which carry the relevant information and computation. Consequently, it is believed that, at any moment in time, the current contents of a mind are determined by the collective states of the neuron membranes i.e., by their momentary voltages [1, 2, 3, 4]. Thus, these two mechanisms, synaptic connections and voltages, have been relied on as the key explanatory tools for all our attempts to understand how a mind may emerge within the brain. This paradigm is generally known as connectionism [1].

However, connectionism did not yet produce a satisfactory explanation of how the mental emerges from the physical. A number of open problems remains. These include unrealistic connectivity patterns [5], inability to work with symbols [6], lack of feed-forward architecture in the brain [7], global workspace for consciousness [8], and others [9]. The list also includes the scaling problem, which will be described next. As a result, the explanatory gap between the mind and the brain remains wide open [10, 8, 11, 12]. This situation raises a question of whether the classical paradigm should be challenged, and an alternative should be proposed—a paradigm that can offer a fresh set of ideas on how to explain the mind within the brain.

## 2. The scaling problem of intelligence

The present argument starts with a notion that animal-level and a human-level intelligence can only be reached by networks able to scale their intelligence well and moreover, classical connectionist networks do not scale intelligence satisfactorily. Recent studies have elucidated the scaling limitations of connectionist networks: Investigations of the best performing connectionist networks, known as deep learning systems, revealed that advances in their intelligence can be only achieved if accompanied with an excessive increase in resources. Across all studies, these demands have been found to grow with a power law [13, 14, 15]. To double the number of objects recognized or to reduce the error of classification by half, an equation of a form $r = i^a$ applies to describe the needed increase in the model size or the size of the training data set ($r$ is the resources needed, $i$ is intelligence and $a$ is the exponent). The value of exponent $a$ is always well above 1 (Figure 1A). This means that the demands on resources explode as the intelligence grows. One study estimated the exponent $a$ to be ~9 for computer vision [14]; this implies that doubling the intelligence of the state-of-the-art deep learning networks for vision (e.g., doubling the numbers of objects recognized without sacrificing accuracy), requires some $2^9$ = ~500-fold increase in resources. Another study implied an exponent of ~13 for language models (~8000-fold increase in resources for doubling the intelligence; Figure 1A) [15].

A question is whether the connections of the brain, which are different from those in deep learning models, can solve the scaling problem. This is unlikely. First, connectionist models of the brain have been around for a while (e.g., [2, 16, 17]); if they were performing better than deep learning, this would have been noticed and would have been already used in AI technology. Second, computational neuroscience has never been able to show theoretical models that deal with anything but toy problems: To the best of the author's knowledge, no theoretical model of the brain has ever been able to cope with a real-life task. This suggests a generally poor scaling properties of connectionist systems. In addition, in Supplementary Materials there is an argument based on generalized logical XOR functions explaining why connectionism cannot scale effectively (applicable to both deep learning or biological networks).

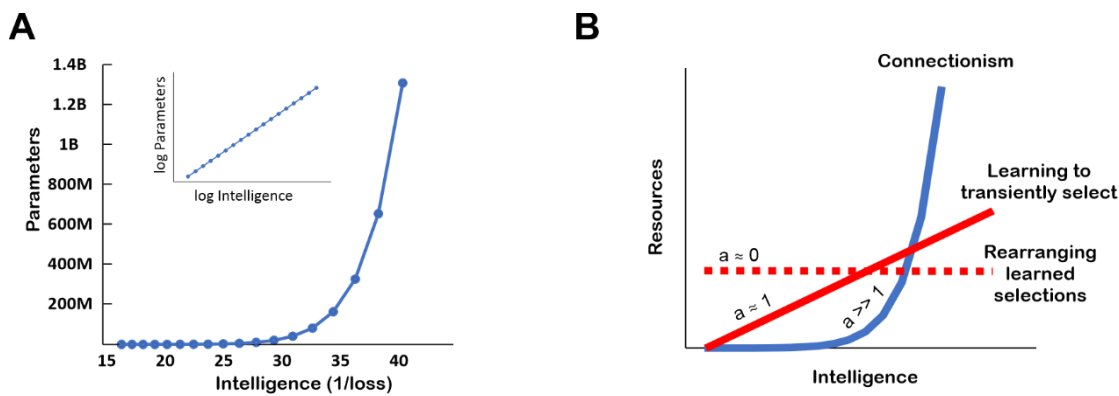

**Figure 1.** The scaling problem in connectionist systems vs. in biological networks. **A)** A power law relation between the intelligence of a transformer model for language processing (intelligence defined as 1/loss) and the number of parameters in the model. Adapted from [15]. Inset: the same graph shown in a log-log plot, the straight line indicating a power law. **B)** Demands on resources for various approaches to intelligence. Blue: In connectionism the demands on resources grow with power law where the exponent $a$ is much larger than one. In biology, the resources grow linearly for learning (solid red line) and remain unchanged for recombining the learned pieces of knowledge ($a$ = 0; dashed red line).

These conclusions stand in contrast to the excellent scaling properties of biological brains. When comparing the brain sizes of different species, it is evident that only an increase in the brain size by a factor of 200 is sufficient to rise the intelligence from the level of a mouse to that of a human. Similarly, the factor of 4 distinguishes the sizes of the chimpanzee brain and of a human. Deep learning models cannot do much with such small increases in the number of parameters, as they would allow for an increase in intelligence at most by 80% (for a 200-fold increase) and by 20% (4-fold increase). If one needs to increase the intelligence of deep learning systems by for example a factor of 1000, which would be a more realistic expectation when comparing a human to a mouse brain, this would require at least a $1000^a = 10^{27}$ if not a $10^{39}$-fold increase in the brain size, which would require a brain of the size of a celestial body.

Even more excessive demands for resources emerge when comparing humans and machines. The number of visual objects that a state-of-the-art connectionist machine may distinguish reliably is in the order of $10^3$. In contrast, biological brains have the capability to effortlessly understand and accurately perceive visual objects created by novel arrangements of elements forming constellations never seen before. An assessment of the total variety of situations that can be correctly perceived by an adult human lead to a lower-bound estimate of $10^{48}$ and the upper bound of $10^{64}$, all without having to resort to new learning [18]. By extrapolating the power laws ruling connectionist machines, the discrepancies remain so vast (in the order of $10^{45}$ to $10^{61}$) that to catch up with the human brain, the power law growth of the state-of-the-art deep learning systems would require them to exceed the size of the known universe.

Thus, it is a must for biological systems to scale their intelligence better than what connectionism can offer. Otherwise, we could not survive in real-world environments, which are more complex than any of the problems that today's AI technology is able to handle. Biological systems exhibit two intelligence-scaling properties that respectively have power law exponents 1 and 0. First, we acquire new knowledge i.e., we learn, with the exponent of about 1 ($a \approx 1$): The number of facts or skills learned is about a linear function of the time we spent learning them; adding to our repertoire a new fact or a skill takes about as much effort as it took to learn a previous fact or skill. This is not the case with connectionist systems which require exploding amounts of training examples and of the training time ($a >> 1$). This is partly due to the need to retrain together new and old items [19] because otherwise, newly learned items start erasing previously learned items, leading to catastrophic interference [20]. Biological brains do not suffer from this problem.

The second intelligence-scaling property of biological systems is the near-zero exponent resulting from the ability to rearrange previously acquired pieces of knowledge into a new combination. Many problems that would be considered new for connectionist networks are in fact just variations of previous problems for our minds. Hence, these problems can be solved by reusing components of previously learned skills. If an intelligent system is able to disentangle any given situation into the correct familiar components, like arranging mental Lego pieces, and is able to learn these Lego pieces instead of learning a solution to the problem by rote, then a possibility emerges to reuse these Lego pieces to solve new problems by rearranging them into novel combinations. This enables humans to drive cars under varying driving conditions without having to practice millions of examples of such driving conditions – in contrast to the connectionist systems. This also enables animals navigate and survive the varying conditions in complex and hostile environments. This Lego-like decomposition of skills also makes our language productive [21]: We can quickly describe new situations and a listener can quickly understand them, even if these situations have never occurred before: "*an astronaut using its oxygen tank to pump up the tires of the space bicycle*" or "*a dinosaur needs privacy when using a dinosaur toilet*". There is a vast number of possible combinations of Lego elements producing situations that we effortlessly understand; the lower bound of this number has been estimated to $10^{48}$ for human brain [18]. These novel arrangements are for our brains often easy and take so little time that the whole effort is negligible in comparison to the efforts connectionists systems need to undertake to learn to deal with such new situations. The present proposal for a new paradigm is based on the presumption that biological cognition with attention, perception, working memory and so on solves the scaling problem by rearranging the Lego pieces. As a result, cognitive operations implemented by biological systems effectively reduce the power law exponent of to a =1 for learning and $a \approx 0$ for the application of learned facts (Figure 1B).

## 3. An alternative paradigm: Transient selection of subnetworks by metabotropic receptors and G protein-gated ion channels

*3.1. Local, sturdy but transient changes to the network wiring*

The idea proposed here is that, at any moment in time, only a fraction of the entire physical neural network is permeable to neural activity: a large portion of dendritic trees and terminal branches are supposed to normally be closed for traffic (are not effectively connected) and thus, do not participate in the flow of neural activity (or the ability to participate in the flow is severely limited) (Figure 2A, B). It is only that the selection of this sub-network is dynamically changed – as needed.

Therefore, the scaling problem is solved such that, at any given moment in time, the brain works with a network of a tiny size. The activated Lego pieces are the pieces of the network that are open for traffic. For example, if a person is trying to get a thread through a needle, it is the best to only activate a network specialized for this task and shut everything else off. Similarly, if one is performing a task of driving a car, another network specialized for this task should be temporarily formed. For example, a network may be formed that acts as a servo mechanism to keep the car in the lane. Similar may apply to any other activity: Any given task is performed by a small network specialized for that task. As a consequence, only a small network undergoes plastic changes alleviating the catastrophic interreference problem and the need for large training data sets. Also, once the subnetwork is activated, a decision complexity of that network reduces only to a small set of alternatives. That way, by continually selecting small networks, the brain remains at all times at the left-most part of the power law function in Figure 1.

Out of a large palette of skills that any person may have, only one (or at most a few) can be active at a time. We choose whether to drive a car or thread a needle or do something else but we cannot do all of those things at the same a time. Cognition is the process of building a new subnetwork i.e., it is a process of choosing. Therefore, decision making is a process of rewiring the brain too. And so is the process of perceiving an object, as much as is the act of directing attention. Generally speaking, every cognitive operation is a process by which the brain is temporarily rewiring itself.

This can be described as *thinking through pathway selection*. For this to work, there are a few preconditions that need be satisfied. The first requirement is that the changes to wiring are decided by the very local networks that are being rewired. No auxiliary (secondary) network is allowed to make such decisions. If an auxiliary network would be responsible for making these decisions, the needed resources would quickly explode (i.e., the power law with a >> 1).

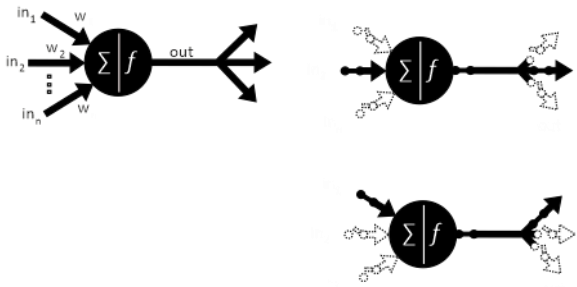
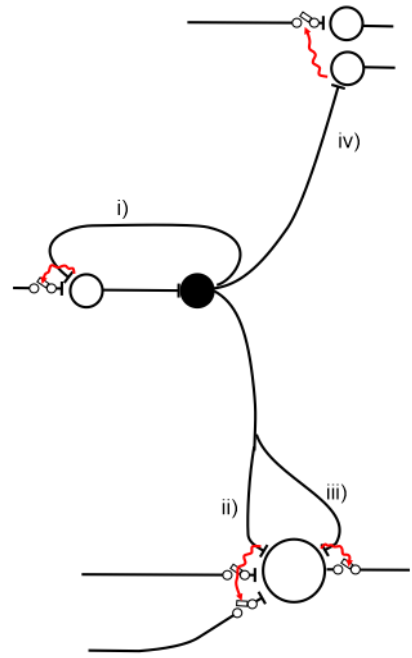
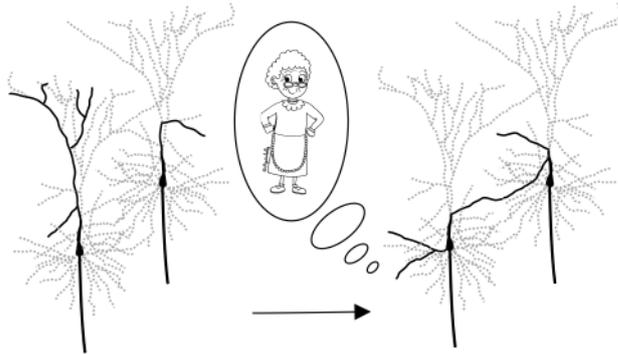
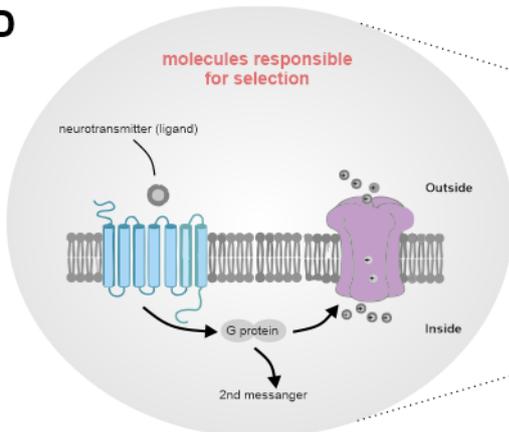
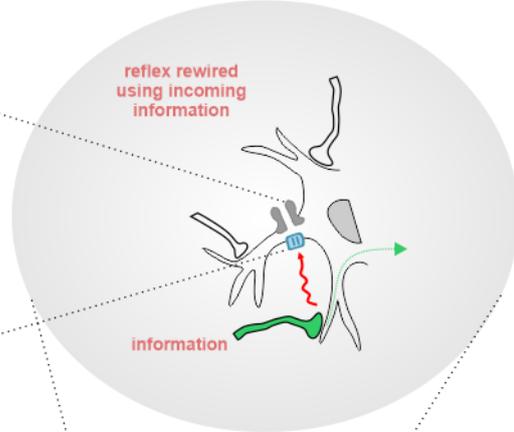
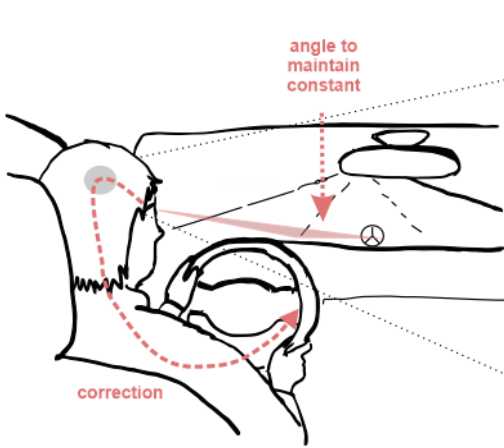
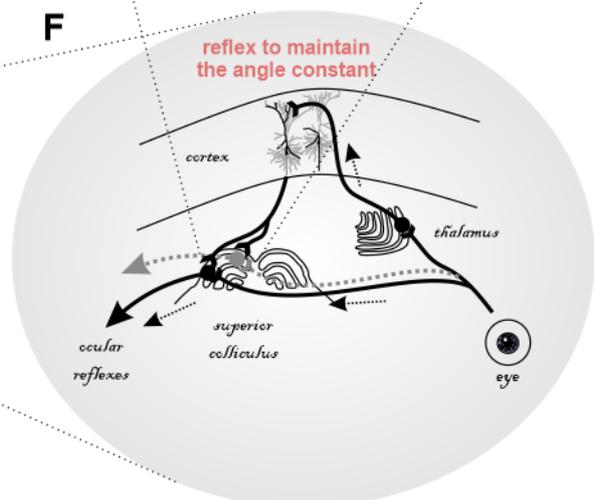

**Figure 2**: **A putative ability of neural networks to transiently select subnetworks. A)** A schematic view on the transient selection hypothesis. i) The classical paradigm assumes that connection strengths are fixed across all tasks that a neuron performs. ii) Transient selection proposes that the routing of activity can temporarily change. Neurons are 'smart' enough to use their recent activity as a source of information for temporarily 'rewiring' their dendritic trees and axonal branches. **B)** A cognitive operation such as perception of a grandmother is executed by transiently altering the wiring of a network: To perceive a grandmother is to temporarily activate a subnetwork specific for interactions with grandmothers (a "grandmother network"). Black: wires opened for traffic; gray: wires closed for traffic. **C)** By sending outputs, a neuron not only inhibits or excites its targets but also provides information on the basis of which the target rewires itself. This information can be sent (i) backwards, (ii) laterally, (iii) forward and (iv) possibly also to neurons with which no direct synaptic connection is formed (disjoined interaction). **D)** Metabotropic receptors (MRs) and G protein-gated ion channels (GPGICs) are the membrane proteins proposed to be responsible for the transient selection capabilities of neural networks. An MR detects presence of a ligand in the extracellular space and releases a G protein which the activates a GPGIC; GPGIC opens or closes and by doing so alters the electrical properties of the membrane. **E)** One source of ligands are 'runaway' neurotransmitters resulting from nearby synaptic activity. **F)** As subnetworks are transiently selected, formed are novel temporary sensory-motor loops specific for the current task. **G)** Example of a temporary sensory-motor loop needed to drive a car – to keep the car within a lane by keeping constant a given visual angle making corrections by turning the steering wheel.

The second requirement is that the changes to wiring are sturdy. The new state of some pathways being selected, and others being shut off must remain in place while the subnetwork is being bombarded with new inputs; incoming activity should not be able to 'wash away' the changes made to wiring. Again, sturdiness should not be brought about by introducing an auxiliary network that would for example, apply recurrent activity. Again, such additional networks would produce an explosion in required resources. Consequently, the system would not scale well. In conclusion, everything needs to be implemented locally, the decisions and the maintenance of changes made.

Finally, the third requirement is that the changes are made in a transient manner: each change should be reverted back after some time. Transiency ensures that one can return to previous tasks, thoughts and ideas. No change should cross the point of no return. The duration over which these transient changes should be valid i.e., until they revert back to the initial state, probably varies and can be estimated from the known duration of our sensory memory. The duration of the visual sensory memory (aka, iconic memory) has been estimated to 200 to 300 ms [22, 23] and that of the auditory sensory memory (aka, echoic memory) to up to 10 seconds [24]. These time windows are much longer than what the changes in neuron's hyper- or depolarization can hold, as the altered states of voltages last no longer than several milliseconds and can easily be 'washed away' by incoming inputs.

*3.2. Classical computational mechanisms cannot solve the scaling problem*

Within the classical paradigm, connectionism, numerous mechanisms have been proposed for temporarily rerouting neural activity. Most commonly, these mechanisms rely on inhibition i.e., GABAergic inputs based on Gamma-aminobutyric acid that serves as the primary inhibitory neurotransmitter [25]. However, other proposals have been made including NMDA activity [26], recursive connectivity [27, 28] and gamma oscillations combined with neural synchrony [16, 4, 29, 26]. However, all of those putative mechanisms suffer from the above-described scaling problems, each missing one or more of the above three requirements.

To save space, I will focus here on the limits of inhibition in solving the scaling problem as this is the most popular one, being commonly used in machine learning (by allowing any synaptic weight to assume negative values). There are multiple problems with inhibition. The most obvious one is that inhibition does not have lasting effect on subnetworks. GABAergic effects on ligand-gated ion channels are not sturdy as they wash away quickly. Consequently, to rewire a network during the period of say 200 ms, GABAergic inputs would need to be applied ten of times in a row or so (assuming 20 millisecond relaxation time [30]). Such sustained inputs would in turn require auxiliary networks of some form, which would lead to an explosion in required resources, as discussed above. Consequently, subnetwork selection by inhibition cannot scale well enough to match the true scaling capabilities of biological networks.

Moreover, the number of inhibitory neurons in the cerebral cortex is too small to match the (already poor) scaling capabilities of machine learning. There are not enough inhibitory neurons to provide a sufficiently high resolution of inhibition. In deep learning, every single synaptic weight can be turned into inhibitory (negative) as the need arises. To match that, the brain would need to provide an inhibitory neuron for each synapse. This is not possible because] inhibitory neurons in the cerebral cortex only account for 10%-25% of the neural population in the cerebral cortex [25, 31, 32, 33, 34]. Thus, on average 4 to 10 pyramidal cells need to share one inhibitory neuron. Thus, the ratio is: one inhibitory neuron per roughly 10,000 synapses. Moreover, inhibitory neurons are not mutually independent but activate each other due to being directly connected by electric synapses (based on gap junctions). This reduces even further the resolution of the patterns with which inhibitory neurons can inhibit pyramidal cells. Therefore, if inhibition was used to select subnetworks, the already unsatisfactory scalability of deep learning models would turn into an even worse scalability of real brain networks.

The fact that dendritic trees of biological neurons can perform much richer computations than can an artificial neuron in a deep learning network [35], does not solve that problem either. The computations of a rich dendritic tree are, in essence, equivalent of a small network of simple neurons. Thus, although dendritic computations provide richer local computation, they do not bring the required sturdiness and transitivity of subnetwork selections.

*3.3. The proposed brain mechanisms for selecting subnetworks: Metabotropic receptors and G protein-gated ion channels*

Importantly, physiological mechanisms exist that have the required properties for effective subnetwork selection: for making local, durable and transient changes to wiring. These are the metabotropic receptors (MRs; aka, G protein-coupled receptors) and G protein-gated ion channels (GPGICs) (Figure 2C-E). These membrane proteins are proposed to give rise to subnetwork selection as they can act as gates: they can close and open the membranes for passage of voltage signals (or make this passage more or less difficult). The job of an MR is to detect a condition under which a change should be made, and the job of GPGIC is to make that change happen. An MR detects a presence of a ligand in the extracellular space and this ligand is presumed to serve as a signal for the change. The presence of the ligand is signaled further to the GPGIC by releasing a G protein intracellularly. G protein then activates GPGIC, either directly by binding on the GPGIC, or indirectly via phosphorylation [36, 37, 38]. The activation of a GPGIC alters the electric properties of the membrane [39, 40] and has the capacity to close or open a branch of a dendritic tree [41, 42] and consequently affects the properties of the circuits [41, 42, 43, 44, 45, 46, 47] (Figures 2C, D)(a more detailed example of a contribution of a GPGICs is provided in the Supplementary Materials).

The act of executing a cognitive operation (attending, perceiving, decision making etc.) is the act of changing the hyper/depolarization conduction properties of the membranes by operations of hundreds or thousands of such proteins, not of changing the voltages of the membranes. Electric properties are changed as MRs and GPGICs distributed across the entire network select subnetworks that will allow passage of activity (Figure 2E).

Notably, by each such selection, also a new, unique sensory-motor loop is created (like the mentioned servo loop for keeping a car in the lane; Figures 2F, G). Voltages and synapses do not do much else but closing these simple sensory-motor loops and bringing in sensory inputs to inform further changes in selected subnetworks. All the 'heavy' work of *thinking through pathway selection* is performed by MRs and GPGICs.

The MRs and GPGICs have all the predispositions necessary to learn their combinations in an effective manner. There are at least 50 types of MRs, making it possible to detect virtually any relevant type of ligand [48]). Many of these ligands are neurotransmitters originating from nearby synapses [49, 50, 51, 52, 53, 54, 55]. In addition, there are many types of GPGICs [56], specialized for various types of ions and producing various effects on the membrane's ability to conduct electric signals [36, 37]. MRs and GPGICs can be found in the central [57] and peripheral nervous systems [58]. This all allows enough flexibility to implement a variety of different rules by which MRs and GPGICs select subnetworks and thereby, give rise to cognition—neural electrical activity only serving the supportive role to execute simple sensory-motor loops and bring in sensory information. See Supplementary Materials also for an argument why MRs and GPGICs operate quickly enough to implement cognition.

Critically, GPGICs maintain a new state of the network sturdily and for an extended period of time, several 100s of milliseconds to minutes: when a channel is activated it remains in the new state uninterrupted until it spontaneously returns back [37]. During the activated state, it is not possible by a voltage input to return the state of GPGIC back. Thus, the change made to activity routing cannot be disrupted or washed away. Therefore, MRs and GPGICs detect a need for change locally and implement this change in a sturdy way. This change is nevertheless transient as the state of the membrane returns back after some period of time (Figure 3).

*3.4. Resolving the scaling problem by taking advantage of situations*

For the intelligence of subnetwork selection systems to scale well, there is one more condition to be satisfied, which is a requirement placed on the world in which these networks operate: The world in which transient networks operate must afford *situations*. A situation is when the complexity of the world reduces and only a small subset of the entire knowledge about that world become relevant. If an animal in in an open space vs. a dense forest, the animal finds itself in different situations. Driving a car is a situation. Walking down a street is another situation. Mating, foraging, grooming and playing are situations that animals exchange. In a situation, certain rules apply and countless others do not apply. Only those rules that apply need to be manifested in the subnetwork selected for that situation. All the other rules, applying in the other situations, are irrelevant and thus should be deactivated. If the world would not afford situations, but everything would be relevant at all times, no advantages could be derived from transient subnetwork selection.

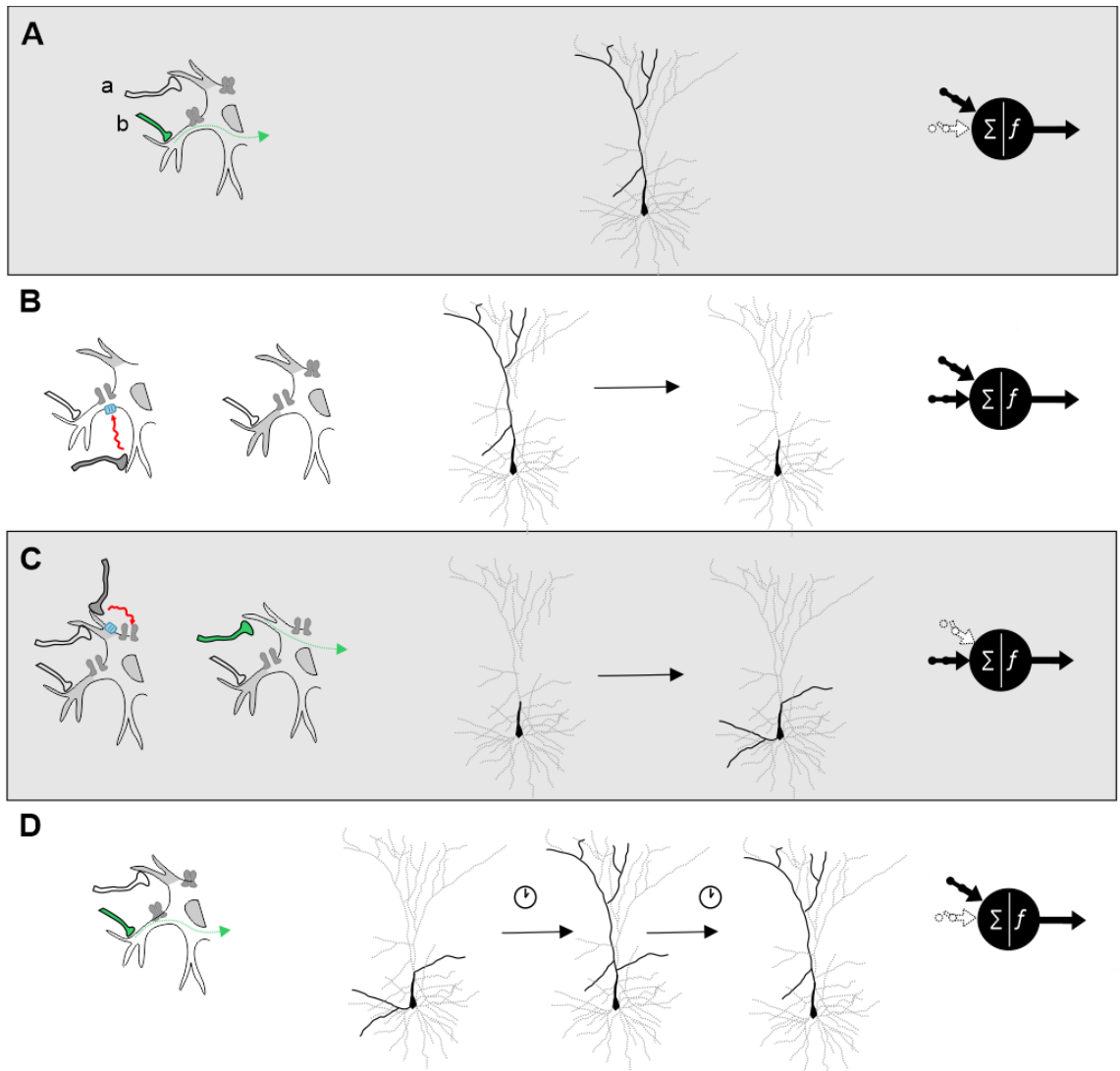

**Figure 3. Sequential accumulation of changes to wiring.** Changes serially accumulate and are eventually released. **A)** A default state of neurons: By default, some wires do not pass activity (*a*), while others do (*b*). **B)** The actions of MRs and GPGICs cause closing of activity passage (or make it more difficult to pass) on one branch of the dendritic tree. **C)** At another branch of the dendritic tree the actions of MRs and GPGICs enable passage through a previously closed part. The changes are being accumulated over events in ii) and iii), the neuron currently having considerably altered wiring properties. These changes may be a part of activating the 'grandmother' network. **D)** Eventually, after some time, GPGICs return to their initial default states. This return occurs without external inputs

As a situation evolves i.e., small changes are made to the situation, also small changes need to be made to the selected subnetwork. For example, a switch from simply keeping a lane to the procedure of changing a lane requires a relatively small change of the selected subnetwork. In contrast, a switch from driving a car to playing badminton is a comparatively larger change in the selected subnetworks. We purposefully organize our lives into manageable situations: brushing the teeth in the morning and then dressing up, changing first a lane and then finding a station on the radio, resolving the issue of one customer before moving to the next customer, taking a break from work to have a lunch, and so on. Animals similarly separate feeding time from foraging time from playing time, and so on.

In a world that would not afford situations, but instead would place demands on one's knowledge at random, a better strategy would be not to rely on transient selection of subnetworks but to directly associate sensory inputs with actions. In that case, one would simply gain no advantages of subnetworks and their formation would consume too much time. In that case, the classical connectionist paradigm would provide a better choice. Connectionist systems have their entire knowledge accessible immediately and at all times. These systems simply recognize patterns and associate them directly to responses. Therefore, transient selection is not always a better paradigm than connectionism. Rather, each approach is associated with a tradeoff. A connectionist system can randomly jump to a new problem without any loss in performance, but the price paid is a need for excessive amounts of resources resulting in poor scaling of intelligence. Transient selection scales much better but it takes extra time to adjust to a newly emerging situations i.e., to 'reprogram' itself. We always take extra time to switch to new tasks [59]. These differences also have implications on how the two types of systems learn. We humans learn best from isolated pieces of knowledge and from stories in which a new piece of information is set into the right context [60, 61]. In contrast, connectionist systems learn best when presented a data set with the entire world knowledge at once or in a fully randomized manner.

*3.5. Brain anatomy supports transient selection of subnetworks*

The new paradigm also requires to re-think how different brain areas cooperate. Classically, it is assumed that cortical areas form a 'pipeline' hierarchy—each area corresponding to a processing stage (like from the brain area V1 to V2 … to IT cortex; Figure 4A). However, in order to scale well, the brain cannot be organized into a serial pipeline, but in a parallel form (Figure 4B). A key problem with a serial pipeline is that the results of computations stored in the states of GPGICs cannot simply be passed onto the next stage in the pipeline. The transitively selected pathways can have effect i.e., can be 'read out', only during the next round of incoming sensory inputs. Therefore, the new paradigm forces us to view the brain as an ultimate iterative device in which its components operate mostly in parallel, not in a pipeline. Sequential processing is achieved only by repeatedly interacting with the environment, the preceding changes to the wiring affecting how the latest stimuli will be processed. This is related to situated (embedded) [62] and enacted [63] cognition. This also means that electrochemical neural activity through alone cannot perform an elaborated form of computation. At each iteration, neural activity only executes simple computations such as closing a sensory-motor loop from Figure 2G.

A consequence is that the new paradigm has a different view on the anatomical connections in the brain. Whereas the connectionist paradigm emphasizes cortico-cortical connections as the primary ones to understand mental operations, the present paradigm emphasizes the connections projecting from subcortical areas to the cortex and again back to sub-cortical areas (Figure 4B). Cortico-cortical projections, albeit important, come only as secondary.

According to the new paradigm, the anatomy of the nervous system can be best understood as repeatedly combining elementary *adaptive* circuits. Elementary adaptive circuits take advantage of both electrochemical neural activity and pathway selection by MRs and GPGICs. Each elementary adaptive circuit has two loops: a primary one based solely on electrochemical activity and a secondary one that additionally engages its MRs and GPGICs (Figure 4C). In the primary loop synaptic inputs arrive, depolarize the cell and output is generated in the form of action potentials (Figures 4C, black). Primary loops can project onto effectors, thereby forming reflexes (e.g., a pseudo unipolar cell and a motor neuron form a primary loop in the spinal cord by executing the patellar reflex). Also, such loops can end at the dendritic tree of neurons anywhere else in the cortex. In this case, the outputs of the primary loop may not activate the next cell i.e., the electrical activity may seem to die out. Nevertheless, the information they provide is not wasted: the synaptic inputs still release neurotransmitters in the extracellular space and these ligands can be detected by nearby MRs [49, 50, 51, 52, 53, 54, 55]. Consequently, information is provided for potential changes to the circuitry to which the primary loop projects (despite a recipient neuron not firing) (e.g., Figure 4C). For example, projections of the pyramidal tract to motor neuron in the spinal cord may adjust the properties of those neurons and thus the reflexes that keep our posture. Projections from primary visual areas to superior colliculus [64] may modulate the sensory-motor loops for following moving objects. And so on.

The secondary loop of the elementary adaptive circuit has to do with its own changes of the states of GPGICs. This is a *rewiring* loop. A circuit does not necessarily passively wait for information on how to change its GPGIC states but may actively ask for this information. The secondary loop sends a second round of outputs as 'corollary' signals. The purpose is not to close an electrochemical loop as in the primary loop, but to inform other parts of the brain about the activity the circuit currently undergoes as to get feedback information on how to accordingly adjust its own responses. This secondary loop is similar to the functions of corollary discharges [65] and efferencer copies [66] of sensory inputs [67]. Here, whenever a signal is sent to a muscle to generate a movement, a copy of the signal is sent to the brain to help interpret whether the subsequent sensory inputs have been generated by the organism's actions or by some other actions in the world. We generalize here the principles underling corollary discharges and efference copies and suggest that these principles generalize across the entire nervous system and across all its circuitry. This is because every bit of the nervous system (with rare exceptions) needs information for adjusting its own responses much like signal to noise distinction for electrolocation in fish [67]. This loop can also be called corollary-for-rewiring loop. For example, a layer 5 pyramidal cell in the cerebral cortex may close the primary loop with thalamus (receiving from and sending back the information to the thalamus), while sending corollary discharges to other nearby pyramidal cells in the same or different mini columns in the cortex. The layer 5 cell receives in turn information from other cortical cells on how to rewire its dendritic tree, which then affects its primary loop with the thalamus in the subsequent iterations (Figure 4B). The brain can take the advantage of dendritic computation, including active dendrites that generate dendritic spikes [35], in a most powerful way when dendritic computation is combined with MRs and GPGICs.

The secondary (corollary-for-rewiring) loop of one elementary adaptive circuit is always connected to one or more primary (electrochemical) loops of other elementary circuits (Figure 4B). This way, the primary and secondary loops across the nervous system become complementary to each other. Neurons seek to close the secondary loops of their own circuits with neurons forming primary loops in other circuits, and vice versa.

This complementary connectivity patterns between electrochemical and corollary-for-rewiring loops are efficient scalers of intelligence. First, due to the complementary relations between elementary adaptive circuits these circuits can be *stacked*: a 'higher-level' circuit that already informs a lower one (or more lower ones) can send its own corollary discharges to receive rewiring instructions from one or more such circuits at even higher levels (Figure 4D). This stack can extend to multiple steps. However, the stack should not be too high: The higher the stack, the slower the information travels down the stack. The backward flow is slow:

While the upward flow can be executed by electrochemical activity and is thus fast, the backward flow involves MRs and GPGICs and a readout of these changes at any step. This readout can be made only in a next iteration of sensory inputs. Thus, the information passes down a stack through multiple sweeps of sensory inputs.

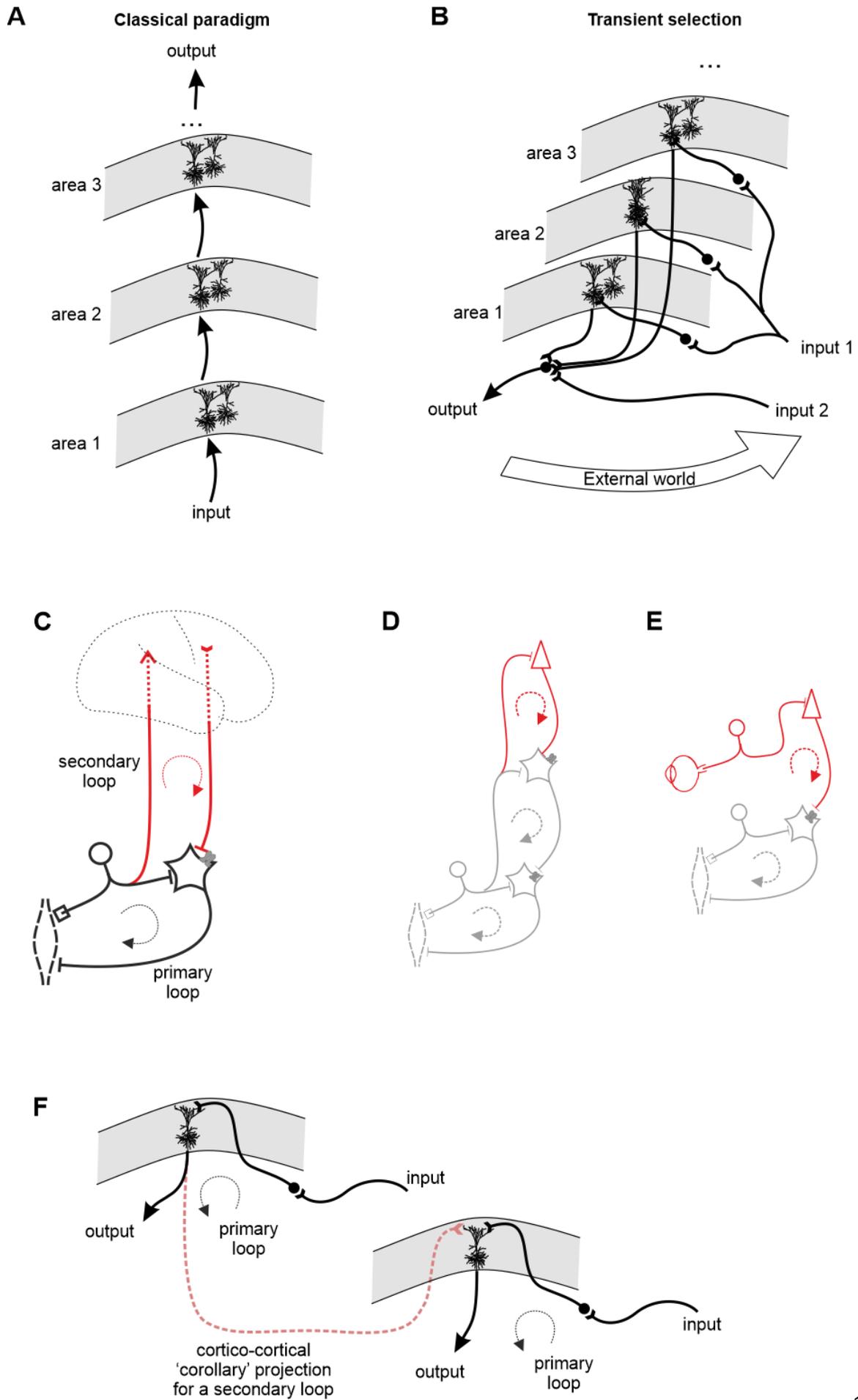

**Figure 4. The wiring patterns that facilitate the new paradigm of subnetwork selection. A)** The sequential (pipeline) computation assumed by the classical paradigm: The computation results from one brain area are passed to the next brain area for further processing. **B)** The results of transient rewiring cannot be passed to the next brain area; hence, the paradigm requires considering parallel operations of across the brain areas, sequential operations being achieved by closing a feedback cycle passing through the external world (i.e., iterations). **C)** Elementary adaptive circuit, consisting of two loops. Only the secondary loop involves rewiring based on MRs and GPGICs. The secondary loop shares similarities to the function of corollary discharges. **D)** Elementary adaptive circuits can be stacked: a circuit that modulates a circuit can be modulated itself but yet another circuit. **E)** Elementary adaptive circuits can crisscross information from other sources. **F)** Cortico-cortical connections are at the top of the stack of elementary adaptive circuits. Cortical areas talk to each other by crisscrossing information for secondary loops: they inform each other how to locally select subnetworks.

The second reason that complementary connectivity from Figure 4C scales intelligence well is the possibility to integrate multimodal information. This is made possible by crisscrossing the inputs and outputs of primary and secondary loops (Figure 4E). The corollary outputs of an elementary circuit of one sensory modality close primary loops such that rewiring information is provided to an elementary circuit in another sensory modality. That way, much of rewiring information comes from other parts of the brain and other sensory modalities. Stacking and crisscrossing of elementary adaptive loops form the brain's global workspace in which transitive changes are made to brain networks in order to organize thoughts, make decisions, perceive objects and so on.

Cortico-cortical connections are at the top of that stacking and crisscrossing hierarchy. They are critical for enabling the cortex as a whole to select the right subnetworks and thus reach the organism's highest levels of intelligence. Without the changes in the state of GPGICs informed through cortico-cortical connections it would probably not be possible to form abstract thought or use math and language. This implies a highly parallel engagement of cortical connections, much different from the serial pipeline of connectionism (Figure 4F).

*3.6 The entire cognition is underpinned by transient rewiring*

The present paradigm is 'radical' in the sense that it does not propose only some aspects of cognition to belong to MRs and GPGICs, like it has been proposed in the past for short-term memory (e.g., [68]), leaving the rest of cognition to the network electrical activity. If cognition was about one half executed by the classical mechanisms and one half by transient rewiring, the scaling problem would not be solved but would only be reduced by 50%. Instead, the present paradigm invites us to consider the *entire* cognition as being executed exclusively by subnetwork selection. This MR and GPGIC-based cognition is riding on top of the stacked loops. All of the various aspects of cognition ranging from attentional filtering to perception, thinking, decision making and insights are suggested to emerge from the functions of MRs and GPGICs, which are particularly rich within the cerebral cortex. Hence, the state of the mind at any moment it time is defined by the state of GPGICs.

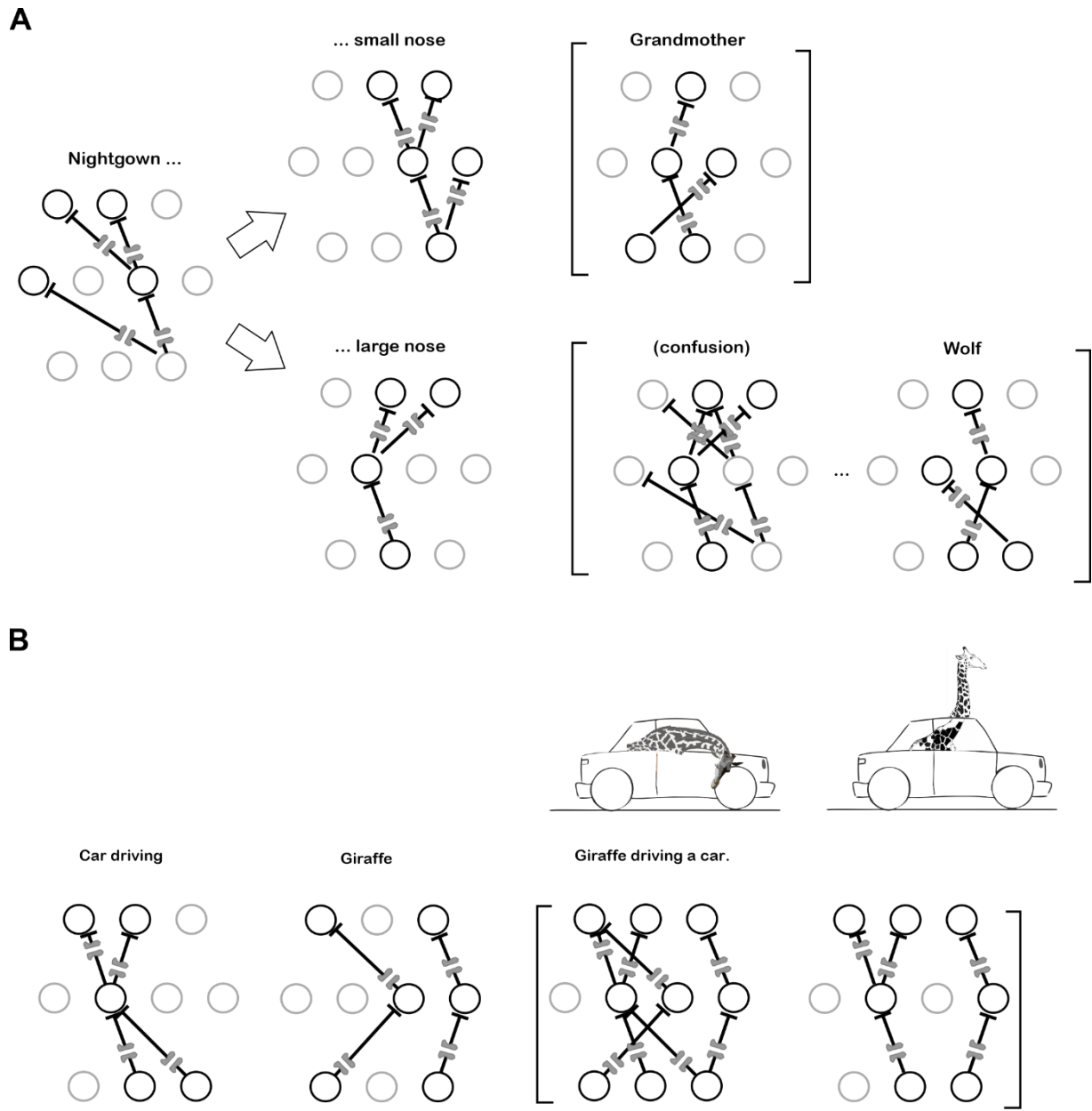

**Figure 5. Schematic illustration of iterative temporal dynamics of transient selections. A)** Cognition emerges from sequential accumulation of selections: Detecting nightgown followed by a small nose settles quickly into a familiar grandmother network; however, nightgown followed by a large nose creates an unsettling combination that demands a more extensive search for the correct familiar network, eventually resulting in the perception of a wolf. **B)** Schematic illustration of a creative composition of existing pieces of knowledge. If one is presented with the idea of 'Giraffe driving a car' for the first time, the subnetworks for 'car driving' and for 'giraffe' interact. Finding a solution may take multiple iteration. The first subnetwork may not be optimal (e.g., inconvenient for the giraffe). In the next iteration the knowledge about cars having sunroofs becomes relevant, allow to find a better way for a giraffe to drive a car.

A canonical scenario of fully engaging all aspects of our cognition may look as follows. Let us assume that a person (e.g., once grandmother) is appearing in a scene: (i) The earliest changes to wiring are made to the sensory systems in order to select potentially interesting inputs (this is attention that affects the direction of eye gaze, head turns and body posture); (ii) These changes direct subsequent sensory inputs which then detect a known object: a grandmother. This may involve multipole iterations. The additional iterations of subnetwork selections prepare the network for interacting with this familiar class of objects (this is perception). Perception includes binding of elementary visual features [69]. Hence, the pre sent paradigm proposes a *binding-by-rewiring* hypothesis. (iii) Further subnetwork selection is needed to choose the appropriate action: Should the grandmother be greeted; by which words? This may require several iterations of selection (here MRs and GPGICs are involved in cognitive operations of thinking and problem solving); (iv) In some cases, the subnetworks that have been so far created by gradually accumulating changes may suddenly undergo a major reorganization. A newly detected relation may be discovered: "It is not a grandmother, it is a wolf dressed into grandmother's nightgown" (this is a cognitive phenomenon of insight) (Figure 5A); (v) Finally, the network may be selected in a way that causes a vigorous action: screaming and running away. In this case MRs and GPGICs select the circuits to boost the energy pumped into the body needed for the fight or flight response (this involves motivation, emotional reactions and decision making).

This iterative nature of learning where previous knowledge assists acquisition of new knowledge, explains why we, unlike connectionist systems, need to learn the basic concepts before learning more advanced skills or knowledge items: counting and addition is learned before calculus, walking before running or playing soccer, and so on. Finally, the Lego pieces that can be flexibly combined are the foundations of our semantics and of our ability to acquire new concepts. Figure 5B illustrates how $a \approx 0$ leads to our ability to solve problems by insights.

This newly created network takes at first several iterations of accumulation to get the subnetwork right; after learning, the same selection is made a lot faster, by fewer iterations of by a single cue (the underlying learning rules are currently not known and still need to be researched [70]). That way, also a new Lego piece is learned which can be used to expand the domain within which the cognition operates, boosting the intelligence of the system as whole even further: now $a \approx 0$ within a one-step broader domain. This is how a person's or animal's intelligence grows iteratively through their lifetime: the existing Lego pieces enable effective learning of new Lego pieces.

Therefore, the present claim is that both reductions in the power law exponent, the $a \approx 1$ for learning and $a \approx 0$ for cognition are achieved through local mechanisms capable of transiently but sturdily selecting subnetworks. They have the capability to learn when and how to make those selections.

## 4. Interpreting electrophysiological experiments through transient rewiring

The argument for transient selection of networks would be incomplete if electrophysiological data would be inconsistent with such ideas and would be hard to interpret in terms of transient selection of subnetworks. Do we observe in the brain activity responses that can be interpreted as a signature of a transient selection of subnetworks? The very first observation once an electrode is placed within a nervous system is that neural responses to stimuli are not constant to constant stimuli. Instead, the responses tend to gradually secede after initial vigorous activity [71, 72]). This phenomenon[, commonly referred to as habituation or fast adaptation [72] is transient [73] and may well be caused by activation of GPGICs and a consequent reduction of the ability of neuron membranes to conduct signals, rather than say by fatigue of a neuron. A related common observation in electrophysiology is that the responses are never identical even if identical stimuli are presented. Instead, neural responses are marked by high variability [74, 75]. This is traditionally considered somewhat mysterious as there is no satisfactory explanation [9]. This variability in responses does not only involve reduced responses i.e., habituation but increases in responses i.e., facilitation. Hence, fatigue cannot possibly explain the variability of brain responses. A possibility is that the underlying network continually changes its properties induced by the operations of MRs and GPGICs.

An even more interesting phenomenon that requires a suitable interpretation is the finding that neural activity following the removal of a stimulus sometimes contains more stimulus-related information (i.e., more stimulus specificity) than the activity during the stimulus presentation. In our studies we found that the information about the identity of flashed stimulus was preserved for at least 500 milliseconds after the stimulus flash was over [76, 77] accompanied by increased firing rates (Figure 6A). This information may not simply reverberate within a classical connectionist networks (as we originally suggested) but may involve transient changes in the reverberating pathways selected by the operations of MRs and GPGICs, prolonging thereby the duration of the memory effect.

Spontaneous activity in intact brains, slices and cultures of neurons [78, 79, 80, 81] is yet another phenomenon that can be explained by transient rewiring of these subnetworks. Spontaneous activity has properties of avalanches [82, 83] and the 'fuel' as well as a trigger for an avalanche may be changes in the membrane properties induced by GPGICs returning back to their deactivated state, lowering the activation thresholds. A triggered avalanche of activity may then activate a set of other GPGICs, which after some time return back to deactivated state and so on – the pattern of activity irregularly repeating.

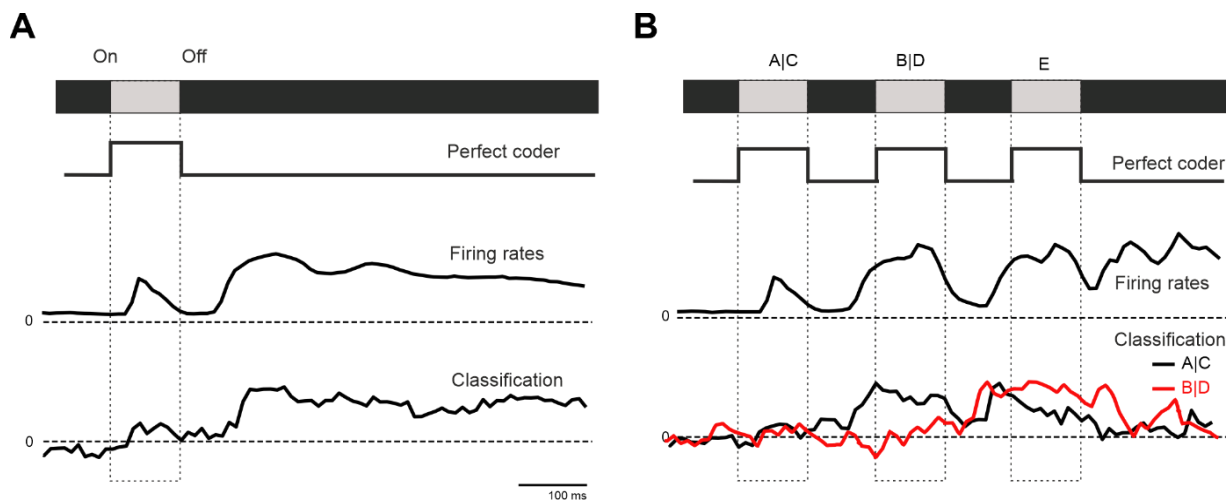

**Figure 6**. **Visual cortex maintains information about recent stimuli**. **A)** Responses to a flashed stimulus by a theoretically perfect coder compared to the response of a population of neurons to such a stimulus in the primary visual cortex (Firing rates), and the stimulus-related information contained within these responses (Classification). Information about the identity of a stimulus is available for several hundreds of milliseconds after the stimulus has been removed. **B)** The same as A) but for a sequence of flashed stimuli. Responses to any stimulus in a sequence contain information about the recently presented stimuli (for example, responses to B|D containing information about A|C). Adapted from [76].

Finally, in one study, we have even observed responses that may present a signature of the process of sequential accumulation of transient selection of pathways [76]. We flashed sequences of three stimuli and the responses to any given stimulus contained not only information about that stimulus but also about the immediately preceding stimulus (Figure 6B). This indicates superposition of information which was accumulated over time. This phenomenon is difficult to explain simply by activity reverberating within a connectionist system. In fact, to the best of author's knowledge, no connectionist model has been made to date capable of accumulating information in such a manner. As described above, the problem is likely in the inability to create sturdy changes which would hold long enough. With the short time constants of neuron depolarizations recursive reverberations likely die out too quickly. Moreover, new inputs likely easily override the existing activity, preventing accumulation of information. Transient selection of pathways based on MRs and GPGICs elegantly explains these results due to their longer time constants.

Further studies will need to apply pharmacological blocking of MRs [84, 85] combined with viral ablation or knockout of GPGICs (e.g., [86, 87]). A prediction is that the blocking these proteins will degrade this ability to superimpose information. Even more elaborate tests can be made by coupling these neural responses with further stimulation in a closed loop [70].

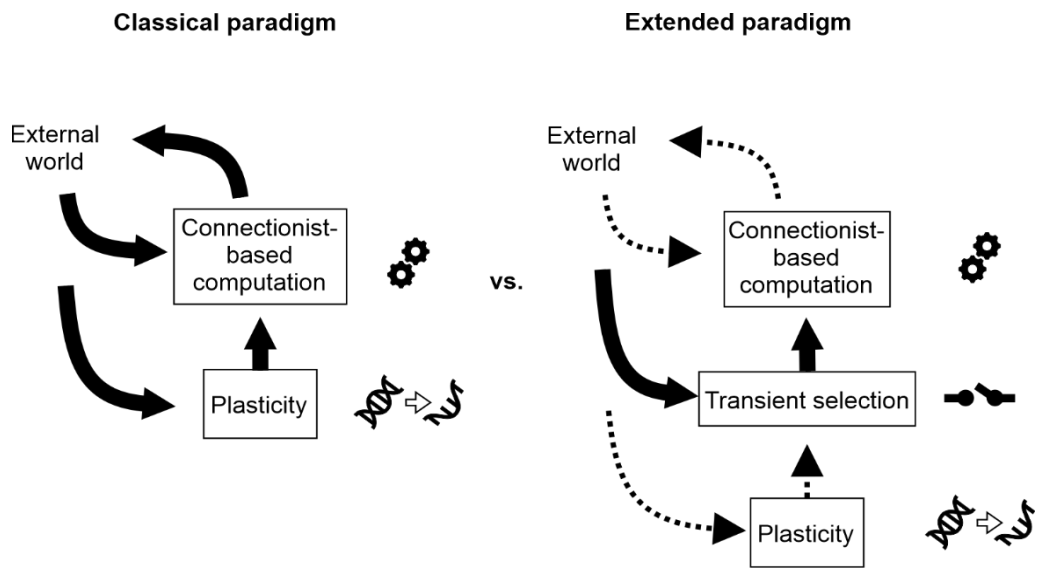

**Figure 7**. **Transient selection paradigm extends the classical paradigm with an additional group of mechanisms.** Left: Two major sets of adaptive mechanisms as presumed by the connectionist paradigm: one set is for computations based on network activity and the other set for plastic changes to that network. Both mechanisms receive feedback from the external world. Right: Transient selection introduces a third set of adaptive mechanisms, transient selection, operating in-between the plasticity and connectionist-based computations and also receiving feedback from the external world.

## 5. Conclusions

Transient selection is a novel approach to explaining the emergence of a mind within a brain and is fundamentally different from the classical connectionist-based paradigm. While the classical paradigm exclusively relies on network connections for storing information and on membrane voltages as the computation medium, the new paradigm extends the capabilities of the nervous system through additional membrane proteins: MRs and GPGICs. MRs detect the conditions for executing 'computation' and the states of GPGICs hold the result of that 'computation'. Consequently, the mental state is defined by the states of GPGICs, not by the state of neuron voltages, as in the classical paradigm. The problem with voltages is that they can too easily be washed away, which then require too large neural networks just to resolve this problem of easy wash-away. There is also a tradeoff, cognition based on subnetwork selection based on MRs and GPGICs can work well only if the surrounding world affords situations. This new paradigm also explains why the brain exhibits its particular connectivity patterns. Biological connectivity patterns are not in a good agreement with connectionism but are in a better agreement with subnetwork selection which requires principles of operations similar to corollary discharges to apply all across the nervous system, including the cerebral cortex. The theory will require much further validation and testing and ideally would involve a visualization method for showing currently active and repressed portions of the synaptic network.

The present paradigm is inspired by the theory of practopoiesis [88]. This theory predicted existence of mechanisms additional to neural (electrochemical) activity and plasticity. This third set of mechanisms was supposed to coordinate the neural activity and yet learn by the means of neural plasticity (Figure 7). As, at the time, a potential physiological mechanism has not yet been identified, these hypothetical mechanisms were named *anapoiesis*, which means 'recreation': Every repeated thought, percept, or a remembered piece of knowledge are being re-creations by the means of selected subnetworks. Subnetworks are formed when the thought is activated, then dismantled to make space for other percepts and thoughts, only to be formed again when the thought or idea is needed again. In the present work, physiological mechanisms for this recreation are proposed: MRs and GPGICs. A computation strategy by first creating the needed networks (i.e., undergoing poiesis) is fundamentally different from the always available and immediately ready-to-go computational mechanisms of connectionism.

Practopoiesis, as a theory, proposes requirements for MRs and GPGICs to successfully undergo learning. This follows from a set of the general principles by which, according to the theory, different levels of organization interact: plasticity interact with MRs and GPGICs (i.e., plasticity for anapoiesis), MRs and GPGICs interact with neural activity (selection of subnetworks). Feedback plays an important role in all cases (Figure 7). Moreover, practopoiesis states that efficient learning can take place only if the learning machinery is rich with knowledge on how and when to learn. In other words, it does not suffice to apply a general all-applicable learning rule as we tend to apply in connectionism (e.g., Hebbian rules, gradient descent). Instead, a significant extent of specialized learning knowledge is required. For example, under a given set of conditions rule A may apply; under another set of conditions rule B may apply, and so on. This learning knowledge (specialized learning rules) are argued to be stored in DNA and to be acquired throughout evolution. Acquisition of additional learning rules make a difference on for example, whether a brain will reach the intelligence of a human or say only of an ape. In terms of machine learning theory, these learning rules can be understood as inductive biases [89]. Furthermore, practopoiesis proposes that these learning rules for establishing MRs and GPGICs are goal oriented; they close a feedback loop (Figure 7) and apply plastic changes until certain targets are reached. An possible experimental design on how to investigate these specialized learning mechanisms and their goals is described in [70].

Finally, the new paradigm based on transient selection of pathways opens a novel approach towards the goal of achieving artificial intelligence (AI) and making it match the capabilities of biological intelligence. Mathematical abstractions of the functionalities provided by MRs and GPGICs can potentially be used to advance machine learning technologies so that they scale as effectively as the human brain.

**Acknowledgments**
The author would like to thank Jaan Aru, Gerald Hahn and Visvanathan Ramesh for feedback on an earlier version of the manuscript. Also, the author would like to thank the two anonymous reviewers whose constructive critique considerably improved the manuscript.
**Conflict of interest:** The author is working on a patent for an AI technology for transient selection of pathways within artificial neural networks.

# References

[1] D. Lloyd, "Consciousness: A connectionist manifesto.," *Minds and Machines,* vol. 5(2), p. 161–1 1995.

[2] "Grossberg, S. (2013). Adaptive Resonance Theory: How a brain learns to consciously attend, learn, and recognize a changing world. Neural networks, 37, 1-47.".

[3] "Friston, K., FitzGerald, T., Rigoli, F., Schwartenbeck, P., & Pezzulo, G. (2017). Active inference: process theory. Neural computation, 29(1), 1-49.".

[4] "Fries, P., Nikolić, D., & Singer, W. (2007). The gamma cycle. Trends in neurosciences, 30(7), 30 316.".

[5] D. Massaro:, "Some criticisms of connectionist models of human performance,," *Journal of Memory and Language,* vol. 27, no. 2, pp. 213-234, 1988.

[6] J. A. Fodor and Z. W. Pylyshyn, "Connectionism and cognitive architecture: A critical analysis.," *Cognition,* vol. 28, no. 1-2, pp. 3-71, 1988.

[7] A. V. I. A. A. &. B. C. Kumar, "Challenges of understanding brain function by selective modulati of neuronal subpopulations," *Trends in Neurosciences,* vol. 36, no. 10, pp. 579-586., 2013.

[8] B. J. Baars, "Global workspace theory of consciousness: toward a cognitive neuroscience of hu experience.," *Progress in brain research,* vol. 150, pp. 45-53, 2005.

[9] J. L. &. S. T. J. (. Van Hemmen, 23 problems in systems neuroscience, Oxford University Press., 2005.

[10] N. Block and R. Stalnaker, "Conceptual analysis, dualism, and the explanatory gap.," *The Philosophical Review,* vol. 108, no. 1, pp. 1-46, 1999.

[11] N. Block, "On a confusion about a function of consciousness.," *Behavioral and brain sciences,* v 18, no. 2, pp. 227-247, 1995.

[12] R. &. K. R. VanRullen, "Deep learning and the global workspace theory," *Trends in Neuroscienc* vol. 44, no. 9, pp. 692-704, 2021.

[13] Y. S. S. H. S. K. K. B.-N. I. G. A. &. K. I. Meir, "Power-law scaling to assist with key challenges in artificial intelligence.," *Scientific reports,* vol. 10, no. 1, pp. 1-7., 2020.

[14] N. C. G. K. L. K. &. M. G. F. Thompson, " The computational limits of deep learning.," arXiv prep arXiv:2007.05558, 2020.

[15] J. M. S. H. T. B. T. B. C. B. C. R. .. &. A. D. Kaplan, " Scaling laws for neural language models.," ar preprint arXiv:2001.08361., 2020.

[16] Rumelhart, D. E., McClelland, J. L., & PDP Research Group. (1988). Parallel distributed processi (Vol. 1, pp. 354-362). New York: IEEE..

[17] "Moca, V. V., Nikolić, D., Singer, W., & Mureşan, R. C. (2014). Membrane resonance enables stable and robust gamma oscillations. Cerebral cortex, 24(1), 119-142.".

[18] D. Nikolić, "Why deep neural nets cannot ever match biological intelligence and what to do about it?.," *International Journal of Automation and Computing,* vol. 14, no. 5, pp. 532-541, 2017.

[19] "Zeng, G., Chen, Y., Cui, B., & Yu, S. (2019). Continual learning of context-dependent processing in neural networks. Nature Machine Intelligence, 1(8), 364-372.".

[20] "McCloskey, M., & Cohen, N. J. (1989). Catastrophic interference in connectionist networks: The sequential learning problem. In Psychology of learning and motivation (Vol. 24, pp. 109-165). Academic Press.".

[21] "Yang, C., Crain, S., Berwick, R. C., Chomsky, N., & Bolhuis, J. J. (2017). The growth of language: Universal Grammar, experience, and principles of computation. Neuroscience & Biobehavioral Reviews, 81, 103-119.".

[22] "Neisser U. (1967). Cognitive Psychology. East Norwalk, CT: Appleton-Century-Crofts".

[23] "Coltheart M. (1980). Iconic memory and visible persistence. Percept. Psychophys. 27, 183–228 10.3758/BF03204258".

[24] "Sams, M., Hari, R., Rif, J., & Knuutila, J. (1993). The human auditory sensory memory trace persists about 10 sec: neuromagnetic evidence. Journal of cognitive neuroscience, 5(3), 363-370.".

[25] "Winer JA, Larue DT. Populations of GABAergic neurons and axons in layer I of rat auditory cortex. Neuroscience. 1989;33(3):499–515. pmid:2636704".

[26] "Lisman, J. E., Fellous, J. M., & Wang, X. J. (1998). A role for NMDA-receptor channels in working memory. Nature neuroscience, 1(4), 273-275.".

[27] "Maass, W., Natschläger, T., & Markram, H. (2002). Real-time computing without stable states: A new framework for neural computation based on perturbations. Neural computation, 14(11), 2531-2560.".

[28] "Donner, R. V., Zou, Y., Donges, J. F., Marwan, N., & Kurths, J. (2010). Ambiguities in recurrence-based complex network representations of time series. Physical Review E, 81(1), 015101.".

[29] "Friston, K. J. (2011). Functional and effective connectivity: a review. Brain Connect. 1, 13–36. doi: 10.1089/brain.2011.0008".

[30] "Branchereau, P., Martin, E., Allain, A. E., Cazenave, W., Supiot, L., Hodeib, F., ... & Cattaert, D. (2019). Relaxation of synaptic inhibitory events as a compensatory mechanism in fetal SOD spinal motor networks. Elife, 8.".

[31] "Peters A, Kara DA. The neuronal composition of area 17 of rat visual cortex. II. The nonpyramidal cells. Journal of Comparative Neurology. 1985;234(2):242–263.".

[32] "Beaulieu C. Numerical data on neocortical neurons in adult rat, with special reference to the GABA population. Brain Research. 1993;609(1–2):284–292.".

[33] "Ouellet L, de Villers-Sidani E. Trajectory of the main GABAergic interneuron populations from early development to old age in the rat primary auditory cortex. Frontiers in neuroanatomy. 2014;8:40.".

[34] "Braitenberg V, Schüz A. Cortex: Statistics and Geometry of Neuronal Connectivity. 2nd ed. Springer-Verlag; 1998.".

[35] "London, M., & Häusser, M. (2005). Dendritic computation. Annu. Rev. Neurosci., 28, 503-532.".

[36] "Burke Jr, K. J., & Bender, K. J. (2019). Modulation of ion channels in the axon: mechanisms and function. Frontiers in cellular neuroscience, 13, 221.".

[37] "Inanobe, A., & Kurachi, Y. (2014). Membrane channels as integrators of G-protein-mediated signaling. Biochimica Et Biophysica Acta (BBA)-Biomembranes, 1838(2), 521-531.".

[38] "Breitwieser, G. E. (1991). G protein-mediated ion channel activation. Hypertension, 17(5), 684-692.".

[39] "Mark, M. D., & Herlitze, S. (2000). G-protein mediated gating of inward-rectifier K+ channels. European Journal of Biochemistry, 267(19), 5830-5836.".

[40] "Lüscher, C., Jan, L. Y., Stoffel, M., Malenka, R. C., & Nicoll, R. A. (1997). G protein-coupled inwardly rectifying K+ channels (GIRKs) mediate postsynaptic but not presynaptic transmitter actions in hippocampal neurons. Neuron, 19(3), 687-695.".

[41] M. Suzuki and M. E. Larkum, "General anesthesia decouples cortical pyramidal neurons.," *Cell,* vol. 180, no. 4, pp. 666-676, 2020.

[42] "Aru, J., Suzuki, M., & Larkum, M. E. (2020). Cellular mechanisms of conscious processing. Trends in Cognitive Sciences, 24(10), 814-825.".

[43] "Chai, C. M., Torkashvand, M., Seyedolmohadesin, M., Park, H., Venkatachalam, V., & Sternberg, P. W. (2022). Interneuron control of C. elegans developmental decision-making. Current Biology.".

[44] "Yates, J. R., Horchar, M. J., Ellis, A. L., Kappesser, J. L., Mbambu, P., Sutphin, T. G., ... & Wright, M. R. (2021). Differential effects of glutamate N-methyl-d-aspartate receptor antagonists on risky choice as assessed in the risky decision task. Ps".

[45] "Jimenez-Martin, J., Potapov, D., Potapov, K., Knöpfel, T., & Empson, R. M. (2021). Cholinergic modulation of sensory processing in awake mouse cortex. Scientific Reports, 11(1), 1-20.".

[46] "Jin, L. E., Wang, M., Galvin, V. C., Lightbourne, T. C., Conn, P. J., Arnsten, A. F., & Paspalas, C. D. (2018). mGluR2 versus mGluR3 metabotropic glutamate receptors in primate dorsolateral prefrontal cortex: postsynaptic mGluR3 strengthen working memo".

[47] "Galvin, V. C., Yang, S. T., Paspalas, C. D., Yang, Y., Jin, L. E., Datta, D., ... & Wang, M. (2020). Muscarinic M1 receptors modulate working memory performance and activity via KCNQ potassium channels in the primate prefrontal cortex. Neuron, 106(4), 649".

[48] "Basith, S., Cui, M., Macalino, S. J., Park, J., Clavio, N. A., Kang, S., & Choi, S. (2018). Exploring G protein-coupled receptors (GPCRs) ligand space via cheminformatics approaches: impact on rational drug design. Frontiers in pharmacology, 9, 128.".

[49] "Barbour, B., & Häusser, M. (1997). Intersynaptic diffusion of neurotransmitter. Trends in neurosciences, 20(9), 377-384.".

[50] "Mohan, A., Pendyam, S., Kalivas, P. W., & Nair, S. S. (2011). Molecular diffusion model of neurotransmitter homeostasis around synapses supporting gradients. Neural computation, 23(4), 984-1014.".

[51] "Pendyam, S., Mohan, A., Kalivas, P. W., & Nair, S. S. (2012). Role of perisynaptic parameters in neurotransmitter homeostasis—computational study of a general synapse. Synapse, 66(7), 608-621.".

[52] "Nicholson, C. (2001). Diffusion and related transport mechanisms in brain tissue. Reports on progress in Physics, 64(7), 815.".

[53] "Rusakov, D. A., & Kullmann, D. M. (1998). Extrasynaptic glutamate diffusion in the hippocampus: ultrastructural constraints, uptake, and receptor activation. Journal of Neuroscience, 18(9), 3158-3170.".

[54] "Oláh, S., Füle, M., Komlósi, G., Varga, C., Báldi, R., Barzó, P., & Tamás, G. (2009). Regulation of cortical microcircuits by unitary GABA-mediated volume transmission. Nature, 461(7268), 1278-1281.".

[55] "Vizi, E. S. (2000). Role of high-affinity receptors and membrane transporters in nonsynaptic communication and drug action in the central nervous system. Pharmacological Reviews, 52(1), 63-90.".

[56] "Koyrakh, L., Luján, R., Colón, J., Karschin, C., Kurachi, Y., Karschin, A., & Wickman, K. (2005). Molecular and cellular diversity of neuronal G-protein-gated potassium channels. Journal of Neuroscience, 25(49), 11468-11478.".

[57] "Chen, X., & Johnston, D. (2005). Constitutively active G-protein-gated inwardly rectifying K+ channels in dendrites of hippocampal CA1 pyramidal neurons. Journal of Neuroscience, 25(15), 3787-3792.".

[58] "LoRusso, E., Hickman, J. J., & Guo, X. (2019). Ion channel dysfunction and altered motoneuron excitability in ALS. Neurological disorders & epilepsy journal, 3(2).".

[59] "Monsell, S. (2003). Task switching. Trends in cognitive sciences, 7(3), 134-140.".

[60] "McGurk H., MacDonald J. (1976). "Hearing lips and seeing voices". Nature. 264 (5588): 746–8".

[61] "Nikolić, D. (2010). The brain is a context machine. Review of psychology, 17(1), 33-38.".

[62] "Gibson, J. J. (1977). The theory of affordances. Hilldale, USA, 1(2), 67-82.".

[63] "Maturana, H. R., & Varela, F. J. (1987). The tree of knowledge: The biological roots of human understanding. New Science Library/Shambhala Publications.".

[64] "W Fries, Cortical projections to the superior colliculus in the macaque monkey: a retrograde study using horseradish peroxidase, J Comp Neurol. 1984 Nov 20;230(1):55-76. doi: 10.1002/cne.902300106.".

[65] "Sperry, R. W. (1950). Neural basis of the spontaneous optokinetic response produced by visual inversion. Journal of comparative and physiological psychology, 43(6), 482.".

[66] "von Holst, E., & Mittelstaedt, H. (1950). Das reafferenzprinzip. Naturwissenschaften, 37(20), 464-476.".

[67] "Fukutomi, M., & Carlson, B. A. (2020). A history of corollary discharge: contributions of mormyrid weakly electric fish. Frontiers in integrative neuroscience, 14, 42.".

[68] "Goelet, P., Castellucci, V. F., Schacher, S., & Kandel, E. R. (1986). The long and the short of long–term memory—a molecular framework. Nature, 322(6078), 419-422.".

[69] "Treisman, A. M., & Gelade, G. (1980). A feature-integration theory of attention. Cognitive psychology, 12(1), 97-136.".

[70] D. Nikolić, "Testing the theory of practopoiesis using closed loops.," in *Closed Loop Neuroscience*, 2016, pp. 53-65.

[71] "Benda, J. (2021). Neural adaptation. Current Biology, 31(3), R110-R116.".

[72] "Dean, I., Robinson, B. L., Harper, N. S., & McAlpine, D. (2008). Rapid neural adaptation to sound level statistics. Journal of Neuroscience, 28(25), 6430-6438.".

[73] "Sun, W., & Dan, Y. (2009). Layer-specific network oscillation and spatiotemporal receptive field in the visual cortex. Proceedings of the National Academy of Sciences, 106(42), 17986-17991.".

[74] "Churchland, M. M., Yu, B. M., Cunningham, J. P., Sugrue, L. P., Cohen, M. R., Corrado, G. S., ... & Shenoy, K. V. (2010). Stimulus onset quenches neural variability: a widespread cortical phenomenon. Nature neuroscience, 13(3), 369-378.".

[75] "Dinstein, I., Heeger, D. J., & Behrmann, M. (2015). Neural variability: friend or foe?. Trends in cognitive sciences, 19(6), 322-328.".

[76] D. H. S. S. W. &. M. W. Nikolić, "Distributed fading memory for stimulus properties in the primary visual cortex.," *PLoS Biology,* vol. 7, no. 12, p. e1000260., 2009.

[77] A. L. C. F. P. S. W. &. N. D. Lazar, " Visual exposure enhances stimulus encoding and persistence in primary cortex," *Proceedings of the National Academy of Sciences,* vol. 118, no. 43, 2021.

[78] "Mao, B. Q., Hamzei-Sichani, F., Aronov, D., Froemke, R. C., & Yuste, R. (2001). Dynamics of spontaneous activity in neocortical slices. Neuron, 32(5), 883-898.".

[79] Creutzfeldt, Otto Detlev, and Otto Creutzfeldt. Cortex cerebri: performance, structural, and functional organization of the cortex. Oxford University Press, USA, 1995..

[80] "Tsodyks, M., Kenet, T., Grinvald, A., & Arieli, A. (1999). Linking spontaneous activity of single cortical neurons and the underlying functional architecture. Science, 286(5446), 1943-1946.".

[81] "Mazzoni, A., Broccard, F. D., Garcia-Perez, E., Bonifazi, P., Ruaro, M. E., & Torre, V. (2007). On the dynamics of the spontaneous activity in neuronal networks. PloS one, 2(5), e439.".

[82] "Hahn, G., Petermann, T., Havenith, M. N., Yu, S., Singer, W., Plenz, D., & Nikolić, D. (2010). Neuronal avalanches in spontaneous activity in vivo. Journal of neurophysiology, 104(6), 3312-3322.".

[83] "Priesemann, V., Wibral, M., Valderrama, M., Pröpper, R., Le Van Quyen, M., Geisel, T., ... & Munk, M. H. (2014). Spike avalanches in vivo suggest a driven, slightly subcritical brain state. Frontiers in systems neuroscience, 8, 108.".

[84] "Suzuki, M., & Larkum, M. E. (2020). General anesthesia decouples cortical pyramidal neurons. Cell, 180(4), 666-676.".

[85] "Zhang, X. Q., Jiang, H. J., Xu, L., Yang, S. Y., Wang, G. Z., Jiang, H. D., ... & Shen, H. W. (2020). The metabotropic glutamate receptor 2/3 antagonist LY341495 improves working memory in adult mice following juvenile social isolation. Neuropharmacology,".

[86] "Anderson, E. M., Loke, S., Wrucke, B., Engelhardt, A., Demis, S., O'Reilly, K., ... & Hearing, M. C. (2021). Suppression of pyramidal neuron G protein-gated inwardly rectifying K+ channel signaling impairs prelimbic cortical function and underlies stress-".

[87] "1. Lepannetier, S., Gualdani, R., Tempesta, S., Schakman, O., Seghers, F., Kreis, A., ... & Gail-ly, P. (2018). Activation of TRPC1 channel by metabotropic glutamate receptor mGluR5 mod-ulates synaptic plasticity and spatial working memory. Frontiers in c".

[88] D. Nikolić, "Practopoiesis: Or how life fosters a mind.," *Journal of Theoretical Biology,* vol. 373, pp. 40-61., 2015.

[89] "T.M. Mitchell. The need for biases in learning generalizations. Department of Computer Science, Laboratory for Computer Science Research, Rutgers Univ (1980), pp. 184-191".

**Supplementary Materials for Nikolić, D. "Where is the mind within the brain? Transient selection of subnetworks by metabotropic receptors and G protein-gated ion channels"**

*The XOR argument*

A yet another perspective on how MRs and GPGICs empower neural networks can be provided by considering implementations of XOR operations. XOR operations are a part of the real life: "one may spend money on a nice dinner or on a theater play, but not both." or "animals either have four legs or two legs and a pair of wings, but not both", and can be generalized into a multi-bit version called inverters (Supplementary Figure 1A). It has been well established that connectionist systems are able to implement XOR functions. However, there are some limitations that prevent connectionist systems to scale their intelligence for any real-life problems that require XOR operations. First, theoretically, if no recursion is involved, inverters i.e., complex XOR functions, can at best be implemented with linear scaling in the connectionist systems (Supplementary Figure 1B)(this is also known as parity problem). This is a great limitation because this means that, even under theoretically most optimal conditions, connectionist systems cannot reduce the amount of computation to less than $a = 1$. With such scaling it is not possible to distinguish to the order of $10^{48}$ categories. The second limitation makes this problem even worse: the learning mechanisms used for deep neural networks are notoriously incompetent in finding the correct connection patterns needed for implementing XOR mapping functions; the practical scaling properties of connectionist systems are much worse than linear when it comes to XOR. Supplementary Figures 1C and D illustrate the results of a computational experiment in which the smallest possible network sizes were found still able to learn inverters of different depths by using a state-of-the-art learning algorithm. The required sizes of the networks exploded.

Supplementary Figure 1E illustrates how the inverter problem can be solved with $a = 0$ by a recurrent system, which extends the assumptions of connectionism by adding delay lines at the right space (even a recurrent neural network can quickly learn such a task by random guessing of connection weights [1]). This recursive approach also scales effectively. Infinite length inverter can be created. Such delay lines would be resource demanding if implement in the brain simply as long wires; a more effective approach to creating a local memory mechanism of short duration is to place a GPGIC onto the membrane and activate it through an MR (Supplementary Figure 1E, right). Thus, one way of thinking about MRs and GPGICs is as local machinery for storing shot-lasting memories in a non-linear form, effectively applying XOR functions on sequential inputs. If arranged accordingly across the nervous systems (by currently unknown learning rules), this machinery conducts operations sufficient to implement cognition.

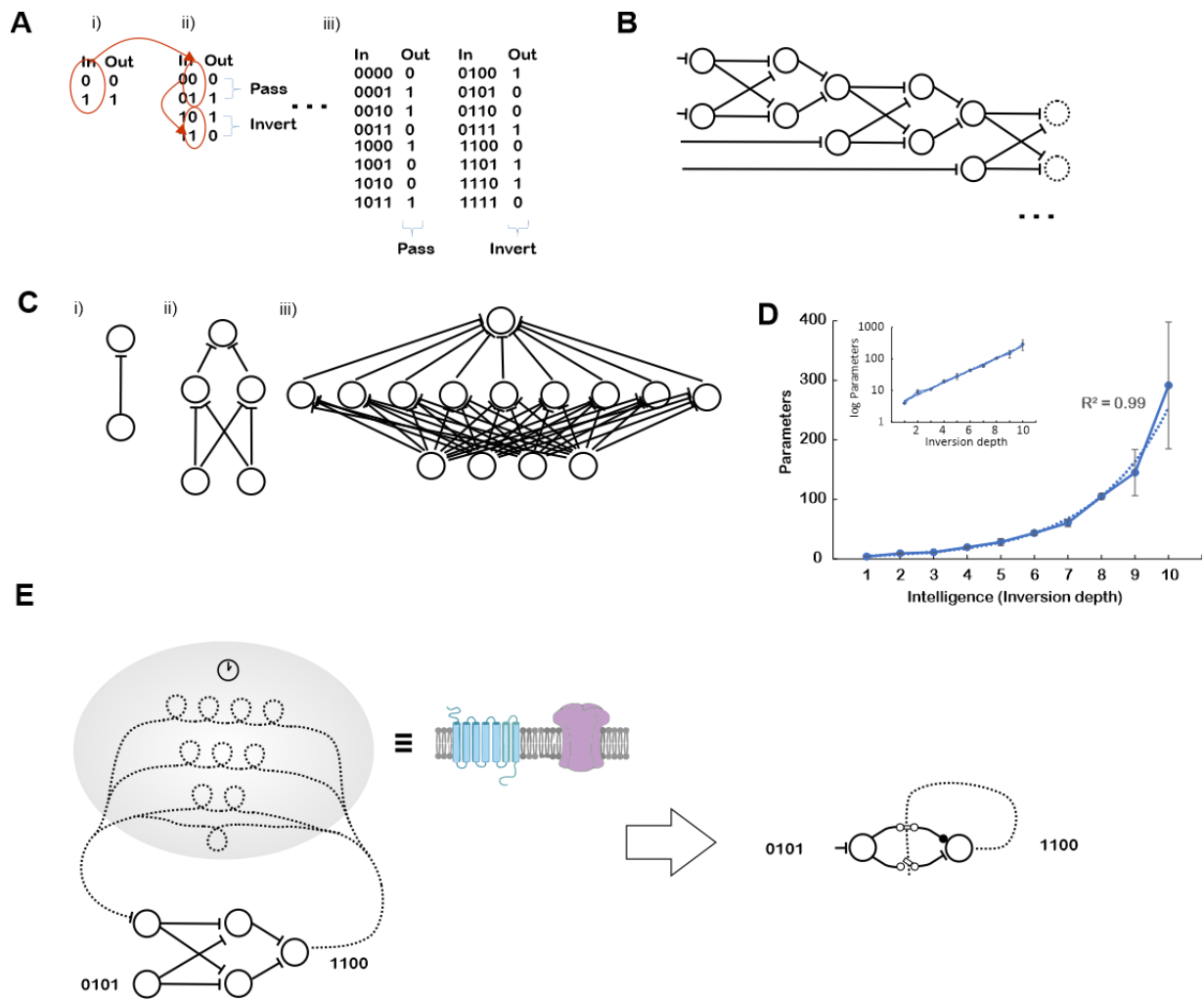

**Supplementary Figure 1. Transient selections for XOR problems. A)** The XOR problem (parity problem) generalized to larger number of bits, known as inverters. i) to iii): inversion depts of 1, 2 and 4 bits. **B)** In theory, connectionist networks can implement the generalized XOR problems (inverters) by linearly stacking XOR circuits. **C)** The results of an experiments: The sizes of networks discovered for inversion depths i) I to iii) in A. **D)** The number of parameters in models for experiments in C) for inversion depths up to 10 bits. Inset: the same data shown in a log-linear graph; the straight line indicating an exponential growth in the demands on resources. Vertical bars: standard deviation across 10 repetitions. **E)** Left: By combining delay lines with a recursive connection, inputs are fed one by one. A single XOR network can implement a generalized XOR of an unlimited inversion depth. Delay lines can be implemented elegantly by MRs and GPGICs. Right: a simple recursive circuit equipped with gates for transient pathway selection produces generalized XOR function.

Inverters are generalizations of the XOR function to an unlimited number of bits. With each additional bit an inverter can be generated by applying the following rule: If the bit is zero, the outputs will be exactly the opposite than if the bit is one. The function poses difficulties when implementing a mapping function of a non-recursive connectionist network because each XOR transformation is non-monotonic. Non-monotonic functions are a subset of non-linear functions and are more difficult to implement than monotonic non-linear functions.

While an inverter (parity) function can easily be constructed even within a non-recursive connectionist network, learning of such a function proves difficult if standard learning methods are use. One needs to devise learning network specialized for that problem (e.g., based on Fourier Transform; [2, 3]. Hence, one has to devise a specialized method for this type of a problem and a unique type of network: Unfortunately, such a 'solution' for learning can be used for real-life problems because these specialized methods cannot learn all the other aspects of the real world. For example, if this rule is to apply in learning to detect a real animal from images "animals either have four legs or two legs and a pair of wings, but not both" (as opposed to a fictional animal) than this rule needs to be applied to learning the visual features of legs and wings, which requires the other type of learning mechanisms such as gradient descent. The reason gradient descent cannot learn inverter functions is that the non-monotonic relationships cannot directly be detected and implemented by the learning rules applied to connectionist networks. Instead, a connectionist network largely needs to stumble upon such an implementation i.e., by chance. The learning rule can only select and sharpen already existing inverter capabilities of networks. As a result, we found that the sizes of randomly initialized networks able to learn inverting functions grow exponentially as a function of the inversion dept (the number of bits in the function), as shown in our study.

We trained network of different sizes and configurations to learn inverter functions of different depths. The depts of the inverter functions ranged from 2 to 10 bits. We used a step-wise increase function, whereby we increased the size of the hidden layer(s) b one neuron. A network would be initialized with random numbers and would be trained until it reached a learning criterion. We used two types of criteria: a low value for loss and accuracy, which needed to be perfect. The low loss criterion required the network to reach a pre-specified value for average loss, which was set to: 0.00001. The exact accuracy criterion required the network to categories the inputs into 1 or 0 output according to the inverter function correctly for all possibly combinations of inputs. The size of the network was increased stepwise; if a network of a given size did not converge, the size of all layers in the network increased by one neuron and the training procedure was repeated. The size of the network was limited to maximum of 150 neurons. To speed up the experiment, an early stop rule was implemented such that a training procedure was declared as not converging if loss did not improve a pre-specified number of steps (patience). Patience varied across experiments and ranged between 25 and 125. The maximum number of epochs allowed for training varied between 4,000 to 200,000. All layers were fully connected and the transfer function was sigmoid. The models had only one output neuron: If output < 0.5 the output was converted to 0, otherwise to 1.

The experiment was performed on GPU GTX 2070 SUPER with 8 GB or RAM and using PyTorch framework version 1.10.1. Learning rules was Adam with the learning rate 0.03. To assess the variability of the network sizes that converged (i.e., learned the inverter function) the procedure was repeated for each inversion depth until the network converged 10 times.

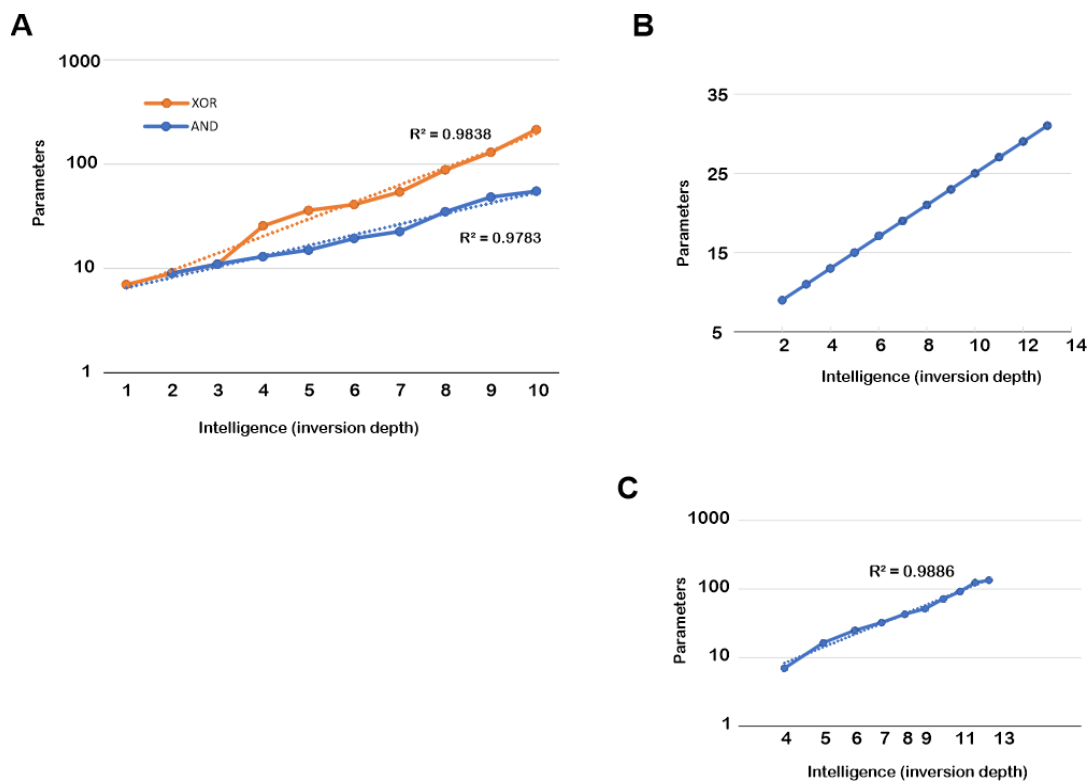

**Supplementary Figure 2**. **A)** The results to the XOR task for the criterion perfect accuracy compared to an AND task with the same criterion. A straight line in a log-linear plot indicates exponential growth in model size. **B)** The same as in B) but for learning logical OR function. A straight line in a linear-linear plot indicates a linear growth in needed resources. **C)** By mixing AND and OR functions, the best fit was obtained by a power law function, as indicated by a straight line in a log-log plot.

The exponential growth in resources reported in the main text was obtained by the low-loss criterion. This result was robust to a change in the criterion: We also obtained an exponential growth in needed resources when perfect accuracy was used as a criterion (Supplementary Figure 2A, orange). We also growth in resources of AND and OR logical operations. Logical AND operation also required an exponential growth in resources, but this growth was much smaller – i.e., it had a much smaller exponent (Supplementary Figure 2A, blue). Therefore, AND is a lot easier operations to learn but it still grows exponentially.

These results implied that AND and XOR logical operations could not explain the power law growth in resources in real-life situations (vision and language). Exponential growth is much more 'aggressive' than power law; exponential growth explodes faster. The difference between the two can be also noted by how the two can be converted into straight lines in plots: to present an exponential function as a straight line, the scale on the ordinate needs to be converted to the opposite of the exponent i.e., to the logarithm (as in Supplementary Figure 2A). In contrast, to visualize a power law function as a straight line, the cases of both abscissa and ordinate need to be presented as a logarithm.) To account for real life's power law growth, we looked at the OR logical operation which turned out to present linear demands on resources (Supplementary Figure 2B). That is, an OR function is the easiest one to learn by deep learning (connectionist) networks. We then propose that the power law function in real life results from a mixture of OR, AND and XOR operations that occur in real life situations. To illustrate that, the results for a 50-50 mixture of logical AND and logical OR are shown in Supplementary Figure 2C: The growth function was fitted best by a power law function.

The code for these calculations can be found here:
https://github.com/dankonikolic/Learning-XOR-AND-OR

*How GPGICs transiently rewire networks*

An example of how the activity of GPGICs affects computations is shown in Supplementary Figure 3. If an inhibitory neuron is stimulated, two types of inhibitory currents are produced in a pyramidal cell of a normal healthy mouse. First there is a short lasting one and followed by a long lasting one (Supplementary Figure 3A). However, if the mouse is genetically altered and the gene for a specific type of GPGIC (the type is GIRK2 in this case) is knocked out. GIRK channels are involved in the depolarization of the membrane [4]. Consequently, the responses of the pyramidal cell to a stimulation of an inhibitory cell show only the fast inhibitory currents while the slow ones are missing (Supplementary Figure 3B). This second, slow component can change the circuit properties for a certain period of time. The stimulus can that way affect how subsequent inputs are being processed. If the GPGIC is missing, inhibition is still present, but only short-lived, the effects being 'forgotten' rapidly.

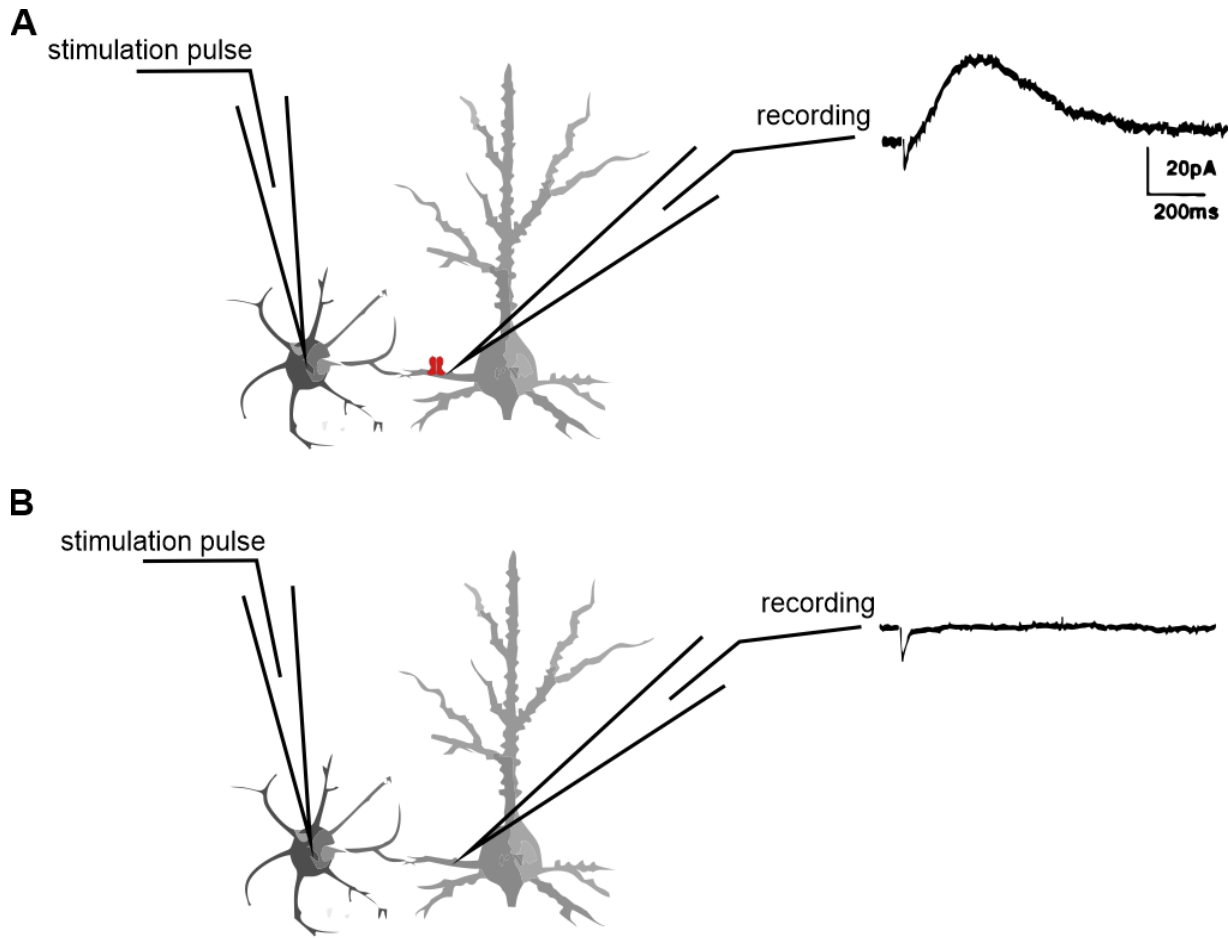

**Supplementary Figure 3.** Inhibitory postsynaptic currents measured form a CA1 pyramidal cell in hippocampus. Stimulation pulse is delivered to an inhibitory neuron. **A)** Responses in a cell of a normal mouse (wild-type). **B)** Responses of a knock-out mouse lacking GIRK2 protein (red only in A). Adapted from ( [5])

Several studies employing similar paradigms as in Supplementary Figure 3 have already reported evidence consistent with the present ideas. By involving anesthetics [6] and blockers of MRs Suzuki and Larkum [7] have shown that preventing MRs from operating, either through anesthetics or blockers, they also reduce the functionality of the dendrites: input signals generate outputs with much more difficulty. Aru et al. [8] interpret the results as MRs being necessary for nervous systems achieving consciousness. The authors propose that MRs and GPGICs switch on the pathways only once to enable signal processing through a dendritic branch. In contrast, the present work proposes that the switching occurs continually "as we think". A more dynamic rule of MRs is shown by Chai et al. [9] who reported an MR based modulation of pathways for making decision of C. elegans under the influence of pheromones. There are also other studies reporting results consistent with the proposed roles of MRs and GPGICs performed on different species ranging from mice to primates [10, 11, 12, 13]. However, admittedly, no study was specifically designed to test the hypothesis that MRs and GPGICs dynamically select subnetwork. Thus, more empirical work is needed.

*Are MRs and GPGICs fast enough to account for cognition?*

Action potentials are fast and so is synaptic transmission; this means that the classical connectionist paradigm is based on fast mechanism: fast voltage gates ion channels for transmission of action potentials and fast ligand gated ion channels for synaptic transmission. The question is than do MRs and GPGICs in combination work fast enough to implement cognition. If the process of activating an MR releasing a G protein and subsequently activating a GPGIC would take minutes, this would be a hard physiological limitation preventing these molecules from taking part in cognitive processes.

Perhaps the most direct evidence about a sufficiently high speed with which these physiological mechanisms can be executed is the processes of phototransduction cascade in rods and cones of retina. When a photon hits rhodopsin, transducin is released which acts as a messenger that binds onto PDE intracellularly. PDE in turn converts cGMP into GMP, reducing intracellular concentration of cGMP eventually closing sodium channels and resulting in the hyperpolarization of the sensory cell [95, 96, 97, 98]. This is a more complex process than the MR-GPGIC cascade proposed to underlie cognition. Yet, the speed of those mechanisms is sufficiently fast to enable our vision. This indicates that the conformational changes to membrane proteins rhodopsin and PDE as well as the diffusion processes of the messengers are fast enough to be employed in cognition. The times of the entire phototransduction cascade should be at most in the order of 10s of milliseconds [16]. Likely, similar speeds apply to MRs and GPGICs on neurons. Moreover, given that a simplest possible form of visual perception i.e., a simplest cognitive operation, cannot occur faster than 100 milliseconds [18, 19], this period should be just about the time needed to pass all the signals through the nervous system and perform one or two iterations of transient rewiring by MRs and GPGICs.

# References

[1] "J. Schmidhuber and S. Hochreiter. Guessing can outperform many long time lag algorithms. Technical Note IDSIA-19-96, IDSIA, 1996".

[2] "Mansour, Y. (1994). Learning Boolean functions via the Fourier transform. In Theoretical advances in neural computation and learning (pp. 391-424). Springer, Boston, MA.".

[3] "Linial, N., Mansour, Y., & Nisan, N. (1993). Constant depth circuits, Fourier transform, and learnability. Journal of the ACM (JACM), 40(3), 607-620.".

[4] "Mark, M. D., & Herlitze, S. (2000). G-protein mediated gating of inward-rectifier K+ channels. European Journal of Biochemistry, 267(19), 5830-5836.".

[5] "Lüscher, C., Jan, L. Y., Stoffel, M., Malenka, R. C., & Nicoll, R. A. (1997). G protein-coupled inwardly rectifying K+ channels (GIRKs) mediate postsynaptic but not presynaptic transmitter actions in hippocampal neurons. Neuron, 19(3), 687-695.".

[6] "Son, Y. (2010). Molecular mechanisms of general anesthesia. Korean journal of anesthesiology, 59(1), 3-8".

[7] M. Suzuki and M. E. Larkum, "General anesthesia decouples cortical pyramidal neurons.," *Cell,* vol. 180, no. 4, pp. 666-676, 2020.

[8] "Aru, J., Suzuki, M., & Larkum, M. E. (2020). Cellular mechanisms of conscious processing. Trends in Cognitive Sciences, 24(10), 814-825.".

[9] "Chai, C. M., Torkashvand, M., Seyedolmohadesin, M., Park, H., Venkatachalam, V., & Sternberg, P. W. (2022). Interneuron control of C. elegans developmental decision-making. Current Biology.".

[10] "Yates, J. R., Horchar, M. J., Ellis, A. L., Kappesser, J. L., Mbambu, P., Sutphin, T. G., ... & Wright, M. R. (2021). Differential effects of glutamate N-methyl-d-aspartate receptor antagonists on risky choice as assessed in the risky decision task. Ps".

[11] "Jimenez-Martin, J., Potapov, D., Potapov, K., Knöpfel, T., & Empson, R. M. (2021). Cholinergic modulation of sensory processing in awake mouse cortex. Scientific Reports, 11(1), 1-20.".

[12] "Jin, L. E., Wang, M., Galvin, V. C., Lightbourne, T. C., Conn, P. J., Arnsten, A. F., & Paspalas, C. D. (2018). mGluR2 versus mGluR3 metabotropic glutamate receptors in primate dorsolateral prefrontal cortex: postsynaptic mGluR3 strengthen working memo".

[13] "Galvin, V. C., Yang, S. T., Paspalas, C. D., Yang, Y., Jin, L. E., Datta, D., ... & Wang, M. (2020). Muscarinic M1 receptors modulate working memory performance and activity via KCNQ potassium channels in the primate prefrontal cortex. Neuron, 106(4), 649".

[14] "Martemyanov, K. A. (2014). G protein signaling in the retina and beyond: the Cogan lecture. Investigative ophthalmology & visual science, 55(12), 8201-8207.".

[15] "Burns, M. E., & Pugh Jr, E. N. (2010). Lessons from photoreceptors: turning off g-protein signaling in living cells. Physiology, 25(2), 72-84.".

[16] "Frolov, R. V., & Ignatova, I. I. (2020). Speed of phototransduction in the microvillus regulates the accuracy and bandwidth of the rhabdomeric photoreceptor. PLoS computational biology, 16(11), e1008427.".

[17] "Pan, G., Tan, J., & Guo, Y. (2019). Modeling and simulation of phototransduction cascade in vertebrate rod photoreceptors. BMC ophthalmology, 19(1), 1-8.".

[18] "Thorpe, S., Fize, D., & Marlot, C. (1996). Speed of processing in the human visual system. nature, 381(6582), 520-522.".

[19] "VanRullen, R., & Thorpe, S. J. (2001). Is it a bird? Is it a plane? Ultra-rapid visual categorisation of natural and artifactual objects. Perception, 30(6), 655-668.".